\documentclass[openacc]{rstransa}

\newcommand{\Z}{{\mathbb{Z}}}
\newcommand{\R}{{\mathbb{R}}}
\newcommand{\C}{{\mathbb{C}}}
\newcommand{\CP}{{\mathbb{C}P}}

\newcommand{\p}{\partial}

\begin{document}

\title{From Quantum Link Models to D-Theory: \\
A Resource Efficient Framework for the 
Quantum Simulation and Computation of Gauge Theories}

\author{Uwe-Jens Wiese}

\address{Albert Einstein Center for Fundamental Physics, 
Institute for Theoretical Physics, University of Bern, Sidlerstrasse 5, 
CH-3012 Bern, Switzerland}

\subject{Quantum simulation, quantum computation, gauge theories}


\corres{Uwe-Jens Wiese \\
\email{wiese@itp.unibe.ch}}

\begin{abstract}

Quantum link models provide an extension of Wilson's lattice gauge theory in
which the link Hilbert space is finite-dimensional and corresponds to a
representation of an embedding algebra. In contrast to Wilson's parallel
transporters, quantum links are intrinsically quantum degrees of freedom. 
In D-theory these discrete variables undergo dimensional reduction, thus
giving rise to asymptotically free theories. In this way $(1+1)$-d $\CP(N-1)$
models emerge by dimensional reduction from $(2+1)$-d $SU(N)$ quantum spin
ladders, the $(2+1)$-d confining $U(1)$ gauge theory emerges from the Abelian
Coulomb phase of a $(3+1)$-d quantum link model, and $(3+1)$-d QCD arises from 
a non-Abelian Coulomb phase of a $(4+1)$-d $SU(3)$ quantum link model, with
chiral quarks arising naturally as domain wall fermions. Thanks to their 
finite-dimensional Hilbert space and their economical mechanism of reaching
the continuum limit by dimensional reduction, quantum link models provide a 
resource efficient framework for the quantum simulation and computation of 
gauge theories.
\end{abstract}

\begin{fmtext}

\section{Introduction}

Gauge theories play a fundamental role in the standard model of particle 
physics. The strong interaction is described by QCD --- the non-Abelian $SU(3)$
gauge theory of quark and gluon fields. The Abelian $U(1)$ gauge theory of QED
is also relevant in atomic, molecular, and condensed matter physics, and in
quantum optics.

\end{fmtext}

\maketitle

Strongly coupled gauge theories confront us with great computational challenges.
In particular, simulations of their real-time evolution or of their behavior
at non-zero fermion density with classical computers are affected by very
severe sign problems \cite{Tro05}. Quantum simulation and computation have
emerged as very promising tools which circumvent the sign problem because they
work directly with quantum hardware and thus naturally incorporate entanglement
and quantum interference. In this way Feynman's vision \cite{Fey82} of 
simulating complicated physical systems by other well-controlled quantum 
systems has become reality \cite{Gre02}. Quantum simulators \cite{Cir12} are 
special purpose quantum computers which are used as digital \cite{Llo96} or 
analog \cite{Jak98} devices, for example, using ultracold atoms in optical 
lattices \cite{Lew12,Blo12}, trapped ions \cite{Bla12}, photons \cite{Asp12}, 
or superconducting circuits on a chip \cite{Hou12}. A digital quantum simulator 
is a precisely controllable many-body system that is programmed to execute a 
sequence of quantum gate operations. The initial state of the simulated system 
is encoded as quantum information, and the real-time evolution is driven 
stroboscopically by a sequence of quantum gates. In an analog quantum 
simulator, on the other hand, the time evolution proceeds continuously. Analog
devices are limited to simpler interactions, but they can be scaled up to 
larger system sizes. 

Implementing gauge theories on quantum hardware is a 
non-trivial challenge \cite{Wie13,Zoh16,Ban20a}. Several analog 
\cite{Buc05,Zoh11,Zoh12,Ban12,Ban13a,Gla14,Mar14,Kas20} as well as 
digital \cite{Mul10,Tag13,Kas20a,Klc20} constructions for gauge theory quantum 
simulators have already been proposed. Experimental realizations of analog or 
digital quantum simulations or computations of lattice gauge theories, some
based upon quantum link models, include 
\cite{Mar16,Ber17,Klc18,Lu19,Sch19,Goe19,Mil20,Yan20,Ata21}. Here we discuss 
quantum link models as a promising resource efficient regularization of Abelian 
and non-Abelian gauge theories. The goal is to provide a pedagogical 
introduction to those aspects of this alternative formulation of gauge theories 
that are most relevant to upcoming quantum simulation or quantum computation 
applications, rather than reviewing this broad subject as a whole. 

\section{Abelian Lattice Gauge Theories in the
Hamiltonian Formulation}

Lattice gauge theories were introduced by Wegner \cite{Weg71} for a $\Z(2)$ 
gauge symmetry and by Wilson for general Abelian or non-Abelian gauge 
symmetries \cite{Wil74}. Here we construct an Abelian $U(1)$ gauge theory in 
the Hamiltonian formulation \cite{Kog75} with the fundamental variables 
residing on the links of a regular spatial lattice. First, we use quantum 
mechanical analog ``particles'' moving around in the $U(1)$ group manifold, 
which is just a circle $S^1$, as the basic building blocks of the theory. The 
resulting link Hilbert space is infinite-dimensional. Then we construct quantum 
link models \cite{Hor81,Orl90,Cha97} by replacing these basic building blocks 
by quantum links, i.e.\ quantum spins endowed with a gauge symmetry, which 
reside in a finite-dimensional link Hilbert space.

\subsection{Analog ``Particles'' Moving in the Group Manifold 
$U(1) = S^1$ }

The basic building blocks of Wilson's lattice gauge theory are group-valued 
parallel transporters associated with the links connecting neighboring lattice 
sites. The link variables of an Abelian $U(1)$ lattice gauge theory are hence 
complex phases $\exp(i \varphi) \in U(1)$. In order to familiarize ourselves 
with these basic variables, we first consider a simple quantum mechanical 
analog, a ``particle'' that is moving in the group manifold $U(1) = S^1$. A 
quantum mechanical particle of mass $M$ that moves on a circle of radius $R$ 
has a moment of inertia $I = M R^2$ and is described by its angular position 
$U = \exp(i \varphi)$, $U^\dagger = \exp(- i \varphi)$, $\varphi \in ]-\pi,\pi]$.
The particle's angular momentum operator plays the role of an electric
field in the gauge theory and is given by $E = - i \partial_\varphi$ (in units 
where $\hbar = 1$). The corresponding commutation relations take the form
\begin{equation}
[E,U] = U \ , \quad [E,U^\dagger] = - U^\dagger \ , \quad [U,U^\dagger] = 0 \ ,
\end{equation}
The kinetic energy operator $T$ as well as its spectrum are given by
\begin{equation}
T = \frac{E^2}{2 I} \ , \quad [T,E] = 0 \ , \quad
T|m\rangle = \frac{m^2}{2 I} |m\rangle \ , \quad
\langle \varphi|m\rangle = \frac{1}{\sqrt{2 \pi}} \exp(i m \varphi) \ , \quad
m \in \Z \ .
\end{equation}
Since the number of eigenstates is infinite, the Hilbert space is
infinite-dimensional.

Let us now consider three ``particles'' moving on $S^1$. We associate the 
``particles'' with the links 12, 23, and 31 that connect the sites 1, 2, 3 of a 
triangle. Their angular momenta turn into the electric fields of a gauge
theory on a triangular lattice $E_{12} = - i \partial_{\varphi_{12}}$,
$E_{23} = - i \partial_{\varphi_{23}}$, $E_{31} = - i \partial_{\varphi_{31}}$. The 
corresponding Hamiltonian contains a specific 3-body interaction
\begin{eqnarray}
H&=&T_{12} + T_{23} + T_{31} + V_{123} = 
\frac{E_{12}^2}{2 I} + \frac{E_{23}^2}{2 I} + \frac{E_{31}^2}{2 I} -
\frac{1}{e^2} \cos(\varphi_1 + \varphi_2 + \varphi_3) \nonumber \\
&=&\frac{E_{12}^2}{2 I} + \frac{E_{23}^2}{2 I} + \frac{E_{31}^2}{2 I} -
\frac{1}{2 e^2}(U_{12} U_{23} U_{31} + U_{31}^\dagger U_{23}^\dagger U_{12}^\dagger) \ .
\end{eqnarray}
Due to the special form of the 3-body force, the Hamiltonian commutes with
the three relative angular momenta of the particles
\begin{equation}
G_1 = E_{12} - E_{31} \ , \quad G_2 = E_{23} - E_{12} \ , \quad 
G_3 = E_{31} - E_{23} \ , \quad [H,G_1] = [H,G_2] = [H,G_3] = 0 \ .
\end{equation}
The relative angular momenta $G_1$, $G_2$, $G_3$ turn into the generators of 
infinitesimal gauge transformations associated with the lattice sites.

\subsection{Many ``Particles'' in  $S^1$ Forming a $U(1)$
Lattice Gauge Theory}

The quantum mechanical analog ``particles'' moving in the group space $U(1)$ 
are used to build a Wilsonian lattice gauge theory. In that case, the 
``particles'' embody parallel transporters $U_{xy} \in U(1)$ associated with 
the links $\langle xy \rangle$ connecting neighboring lattice sites $x$ and $y$.
A $U(1)$ gauge theory on a triangular lattice is then described by the
Hamiltonian
\begin{equation}
H = \frac{e^2}{2} \sum_{\langle xy \rangle} E_{xy}^2 -
\frac{1}{2 e^2} \sum_{\langle xyz \rangle} 
(U_{xy} U_{yz} U_{zx} + U_{zx}^\dagger U_{yz}^\dagger U_{xy}^\dagger) \ .
\label{U1Hamiltonian} 
\end{equation}
Here $\langle xyz \rangle$ denotes a triangular plaquette. We have identified 
the moment of inertia $I = 1/e^2$ as a function of the gauge coupling $e$.

\begin{figure}[tbp] 
\centering \vskip-0.5cm
\includegraphics[width=0.325\textwidth]{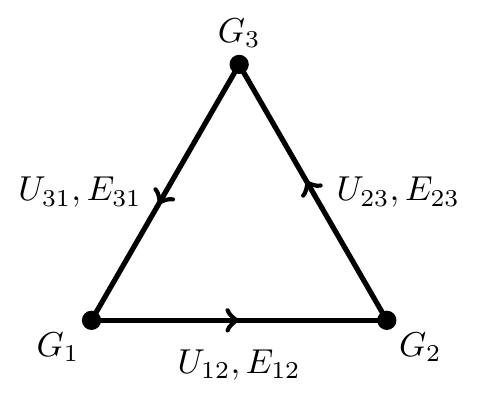} 
\includegraphics[width=0.325\textwidth]{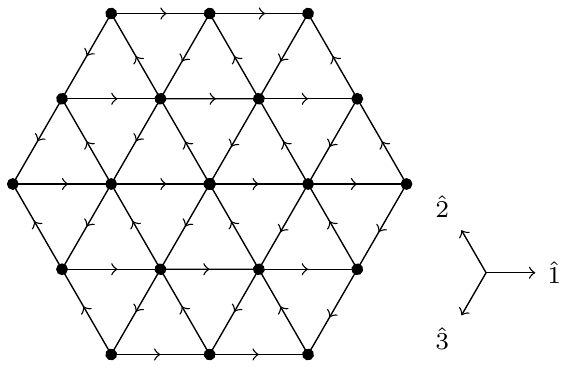}
\includegraphics[width=0.325\textwidth]{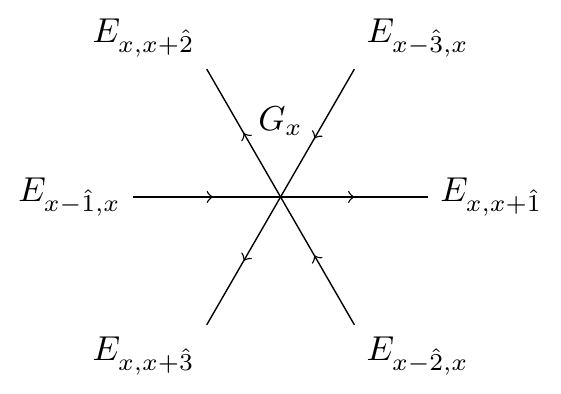}
\caption{\textit{Left: Triangular plaquette with link degrees of freedom
$U_{xy}$, $E_{xy}$, and gauge generators $G_x$ at the sites $x$. Middle:
Triangular lattice with three spatial directions $\hat i$. Right: Gauss law
at $x$ as the lattice divergence of the electric flux.}}
\label{Fig1}
\end{figure}

The structure of this lattice gauge theory is characterized by the 
link-based operator algebra
\begin{eqnarray}
&&[E_l,E_{l'}] = 0 \ , \quad [E_l,U_{l'}] = i \delta_{l l'} U_l \ , \quad
[E_l,U_{l'}^\dagger] = - i \delta_{l l'} U_l^\dagger \ , \nonumber \\
&&[U_l,U_{l'}] = [U_l^\dagger,U_{l'}^\dagger] = [U_l,U_{l'}^\dagger] = 0 \ .
\end{eqnarray}
In particular, the operators $E_l$ or $U_{l'}$, which reside on different links
$l$ and $l'$, commute with each other. As a result of these commutation 
relations, the Hamiltonian commutes with the generators of gauge 
transformations associated with the lattice sites $x$
\begin{equation}
G_x = \sum_i (E_{x,x + \hat i} - E_{x - \hat i,x}) \ , \quad [H,G_x] = 0 \ .
\end{equation}
Here $\hat i$ is the unit-vector pointing in one of the three lattice 
directions, $i \in \{1,2,3\}$, of the triangular lattice (cf.\ Fig.\ref{Fig1}).

It is important to note that gauge symmetries are qualitatively different
from global symmetries. While global symmetries may give rise to degeneracies 
in the physical spectrum, gauge symmetries just reflect a redundancy in the 
description of the physics. Gauge invariance guarantees that the redundancy 
does not affect physical results. This is a consequence of Gauss' law, which 
implies that all physical states $|\Psi\rangle$ must be gauge invariant,
$G_x |\Psi\rangle = 0$. Gauge transformations associated with different sites 
commute with each other, $[G_x,G_y] = \delta_{xy}$, as well as with the 
Hamiltonian. The eigenstates $|\Psi, Q\rangle$, with $Q = \{Q_x\}$, of the 
Hamiltonian can be characterized by the eigenvalues $Q_x \in \Z$ of all gauge
generators, $G_x |\Psi, Q\rangle = Q_x |\Psi, Q\rangle$. Due to Gauss' law, 
the physical Hilbert space is drastically reduced to the states with $Q_x = 0$.
Still, one can assign a physical meaning to the states
$|\Psi, Q\rangle$ with some $Q_x \neq 0$. Those represent a system in the
presence of external static charges $Q_x \in \Z$. As a consequence of the
compact nature of the gauge group $U(1)$, the charges are quantized in integer
units. The canonical quantum statistical partition function for a gauge theory 
is
\begin{equation}
Z_Q = \mbox{Tr}[\exp(- \beta H) P_Q] \ .
\end{equation}
Here $P_Q$ is an operator that projects on the appropriate charge sector. It is 
interesting to investigate a lattice gauge theory in the presence of two
opposite external charges $Q_x = 1$, $Q_y = - 1$, located at different lattice 
sites $x$ and $y$. The potential $V(x-y)$ between the charges is then given by
\begin{equation}
\frac{Z_Q}{Z} = \exp(- \beta V(x-y)) \ , \quad V(x - y) \sim \sigma |x - y| \ .
\end{equation}
Generically, at strong gauge coupling $e$ and at low temperature (large 
$\beta$), lattice gauge theories with a compact gauge group (such as $U(1)$) 
are confining with a linearly rising charge-anti-charge potential that is 
characterized by the string tension $\sigma$. 

\subsection{Quantum Spins as Building Blocks of Abelian 
Quantum Link Models}

We now replace the analog ``particle'' by a quantum spin 
$S \in \{0,\tfrac{1}{2},1, \tfrac{3}{2},\dots\}$, acting in a 
$(2 S + 1)$-dimensional Hilbert space, and obeying the standard commutation 
relations $[S^a,S^b] = i \varepsilon_{abc} S^c$, $S^\pm = S^1 \pm i S^2$,
$[S^3,S^+] = S^+$, $[S^3,S^-] = - S^-$, $[S^+,S^-] = 2 S^3$. This resembles the 
commutation relations $[E,U] = U$, $[E,U^\dagger] = - U^\dagger$ of the analog 
``particle'', if we identify $S^3$ with $E$, $S^+$ with $U$, and $S^-$ with 
$U^\dagger$. However, the relation $[S^+,S^-] = 2 S^3$ does not match 
$[U,U^\dagger] = 0$. This is because the latter only holds in an 
infinite-dimensional Hilbert space. We now introduce the Hamiltonian
\begin{equation}
H = \frac{(S^3)^2}{2 I} \ , \quad [H,S^3] = 0 \ , \quad E_m = \frac{m^2}{2 I} 
\ , \quad m \in \{-S,-S+1,\dots,S-1,S\} \ .
\end{equation}
For large integer spin, $S \in \Z$, its energy spectrum resembles the one of 
the ``particle'' Hilbert space. Interestingly, for half-odd-integer values of
$S$ the quantum spin Hamiltonian provides additional opportunities which lead 
to theories that are inaccessible in the Wilson framework.

\subsection{$U(1)$ Quantum Link Models}

Now we introduce an alternative approach to lattice field theory, which uses 
intrinsically quantum mechanical degrees of freedom --- in this case $U(1)$ 
quantum links --- which are quantum spins endowed with a gauge symmetry. 
Quantum spins reside in a finite-dimensional Hilbert space and are directly 
provided by Nature as a natural candidate for quantum hardware. Quantum spins 
$\tfrac{1}{2}$ embody the concept of a qubit. The simplest $U(1)$ quantum links 
are quantum spins $\tfrac{1}{2}$ residing on the links of a lattice. Quantum 
link models provide a generalization of Wilson's lattice gauge theory. In 
particular, they also provide additional models that are inaccessible in the 
standard Wilson framework. At the same time, the Wilson theory is contained in 
the quantum link framework in the ``classical'' limit $S \rightarrow \infty$. 
Although quantum links can be viewed as discrete quantum variables, they 
naturally lead to Hamiltonians with exact continuous local symmetry. 
Universality, which relies on symmetries, guarantees that the same continuum 
limits can be reached as in the standard Wilson framework of lattice field 
theory.

The Hamiltonian of a $U(1)$ quantum link model on a triangular lattice has the 
same form of eq.(\ref{U1Hamiltonian}) as in the Wilson theory, but the operator 
algebra is modified to
\begin{eqnarray}
&&[E_l,E_{l'}] = 0 \ , \quad [E_l,U_{l'}] = i \delta_{l l'} U_l \ , \quad
[E_l,U_{l'}^\dagger] = - i \delta_{l l'} U_l^\dagger \ , \nonumber \\
&&[U_l,U_{l'}] = [U_l^\dagger,U_{l'}^\dagger] = 0 \ , \quad 
[U_l,U_{l'}^\dagger] = 2 \delta_{l,l'} E_l \ .
\end{eqnarray}
Only the last commutator differs from the Wilson theory, for which 
$[U_l,U_{l'}^\dagger] = 0$. This deviation has no effect on the essential 
commutation relation $[H,G_x] = 0$, because 
$G_x = \sum_i (E_{x,x + \hat i} - E_{x - \hat i,x})$ does not depend on $U_l$ or 
$U_l^\dagger$. Consequently, we have constructed an Abelian gauge theory 
with exact $U(1)$ gauge symmetry from discrete quantum link variables that
reside in a finite-dimensional Hilbert space.  

Quantum simulator constructions for $U(1)$ quantum link models with dynamical 
fermions have used, for example, ultracold Bose-Fermi mixtures in optical 
superlattices \cite{Ban12}, while constructions without fermions have been 
based on Rydberg atoms in optical lattices \cite{Gla14} or on superconducting 
quantum circuits \cite{Mar14}. Numerous different aspects of $(2+1)$-d $U(1)$ 
quantum link models have been investigated in \cite{Sha04,Ban13,Ban14,Car17,
Hua19,Tsc19,Bro19,Fel20,Car20,Cel20,Luo20,Sur20,Kan20,Hal20,Ban20,Dam21,Zac21}.

\subsection{The $S = \tfrac{1}{2}$ Quantum Link Model on a 
Triangular Lattice}

Let us consider the $U(1)$ quantum link model on a triangular lattice with the
smallest possible 2-dimensional link Hilbert space corresponding to 
$S = \tfrac{1}{2}$. Since then $(S^3)^2 = \frac{1}{4}$, the electric field term
in the Hamiltonian is a trivial constant, which can be omitted such that
\begin{equation}
H = - J \sum_{\langle xyz \rangle} 
(U_{\langle xyz \rangle} + U_{\langle xyz \rangle}^\dagger + 
\lambda (U_{\langle xyz \rangle} + U_{\langle xyz \rangle}^\dagger)^2) \ , \quad 
U_{\langle xyz \rangle} = U_{xy} U_{yz} U_{zx} \ , \quad J = \frac{1}{2 e^2} \ .
\label{U1QLMHamiltonian}
\end{equation}
We have added a term proportional to $\lambda$. This term is analogous to the
Rokhsar-Kivelson term \cite{Rok88} in the quantum dimer models of condensed 
matter physics, which are considered in the context of high-temperature 
superconductivity. The model of eq.(\ref{U1QLMHamiltonian}) has a rich 
confining dynamics that is not accessible in the Wilson framework. In 
particular, it has ``nematic'' confined phases for which the discrete lattice 
rotation invariance is spontaneously broken \cite{Ban21}. 

By an exact duality transformation one can construct height variables 
associated with the hexagonal lattice that is dual to the original triangular 
lattice. The dual lattice consists of two sublattices $A$ and $B$. The height 
variables on sublattice $A$ reside at the center $\widetilde x$ of a triangle 
and take values $h_{\widetilde x}^A \in \{0,1\}$. The height variables on 
sublattice $B$, on the other hand, take the values 
$h_{\widetilde x}^B \in \{- \tfrac{1}{2},\tfrac{1}{2}\}$. The electric flux
connecting the sites $\widetilde x = x + \frac{1}{3}(\hat i - \hat j)$ and
$\widetilde x' = x + \frac{1}{3}(\hat i - \hat k)$, where 
$j = (i-1) \mbox{mod} \, 3$ and $k = (i+1) \mbox{mod} \, 3$, is given by
$E_{x,x+\hat i} = \left(h^A_{\widetilde x} - h^B_{\widetilde x'}\right) \mbox{mod} \, 2
= \pm \tfrac{1}{2}$. This relation guarantees that the Gauss law is satisfied 
modulo 2. The full Gauss law results from an additional constraint on the 
height variables.

The phases of the model are distinguished by two sublattice order parameters
\begin{equation}
M_A = \frac{2}{L^2}
\sum_{\widetilde x \in A} \left(h_{\widetilde x}^A - \frac12\right), \quad
M_B = \frac{2}{L^2} \sum_{\widetilde x \in B} h_{\widetilde x}^B, \quad
M_A, M_B \in [-1,1] \ .
\label{orderparameters}
\end{equation}
\begin{figure}[tbp] 
\centering \vskip-0.5cm
\includegraphics[width=0.325\textwidth]{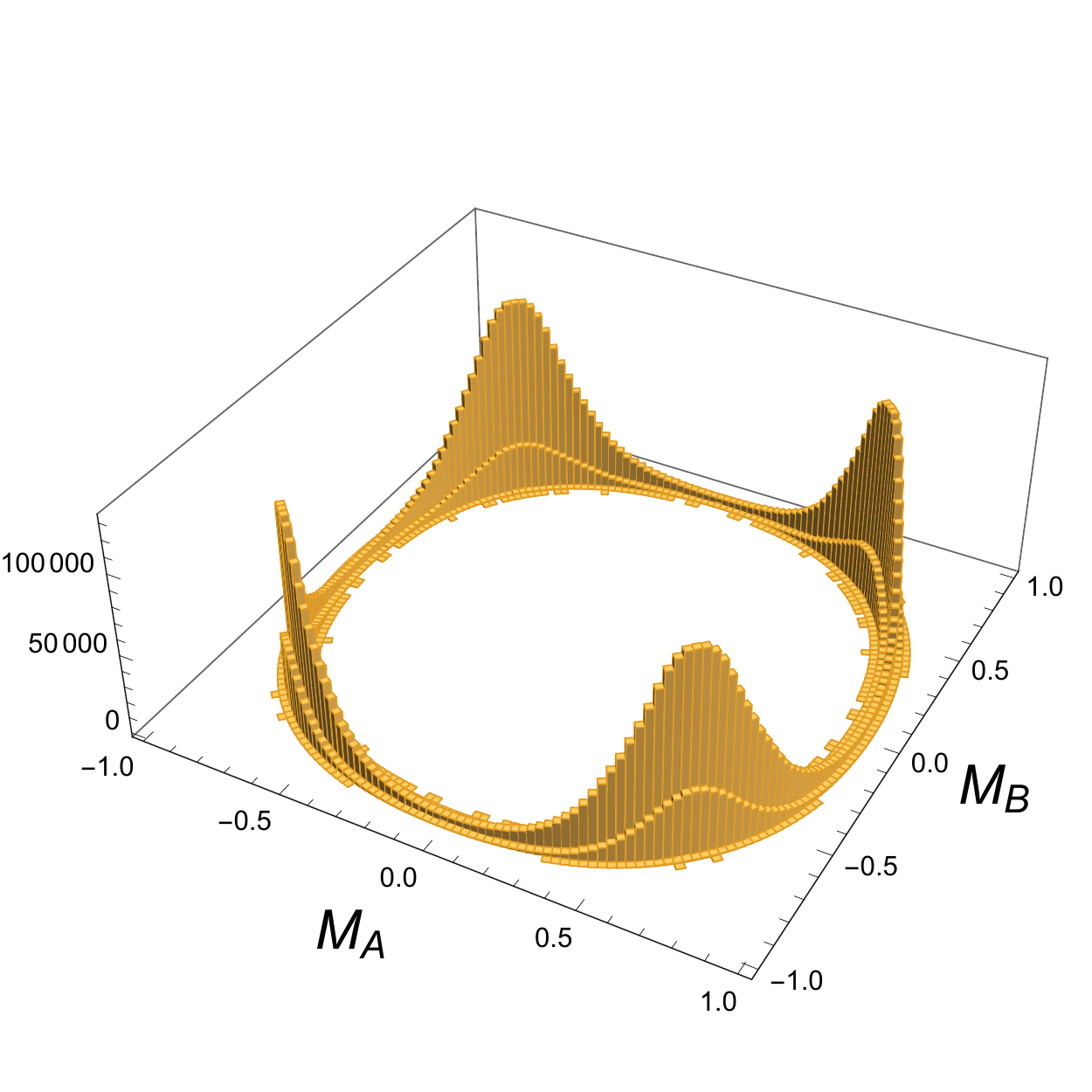} 
\includegraphics[width=0.325\textwidth]{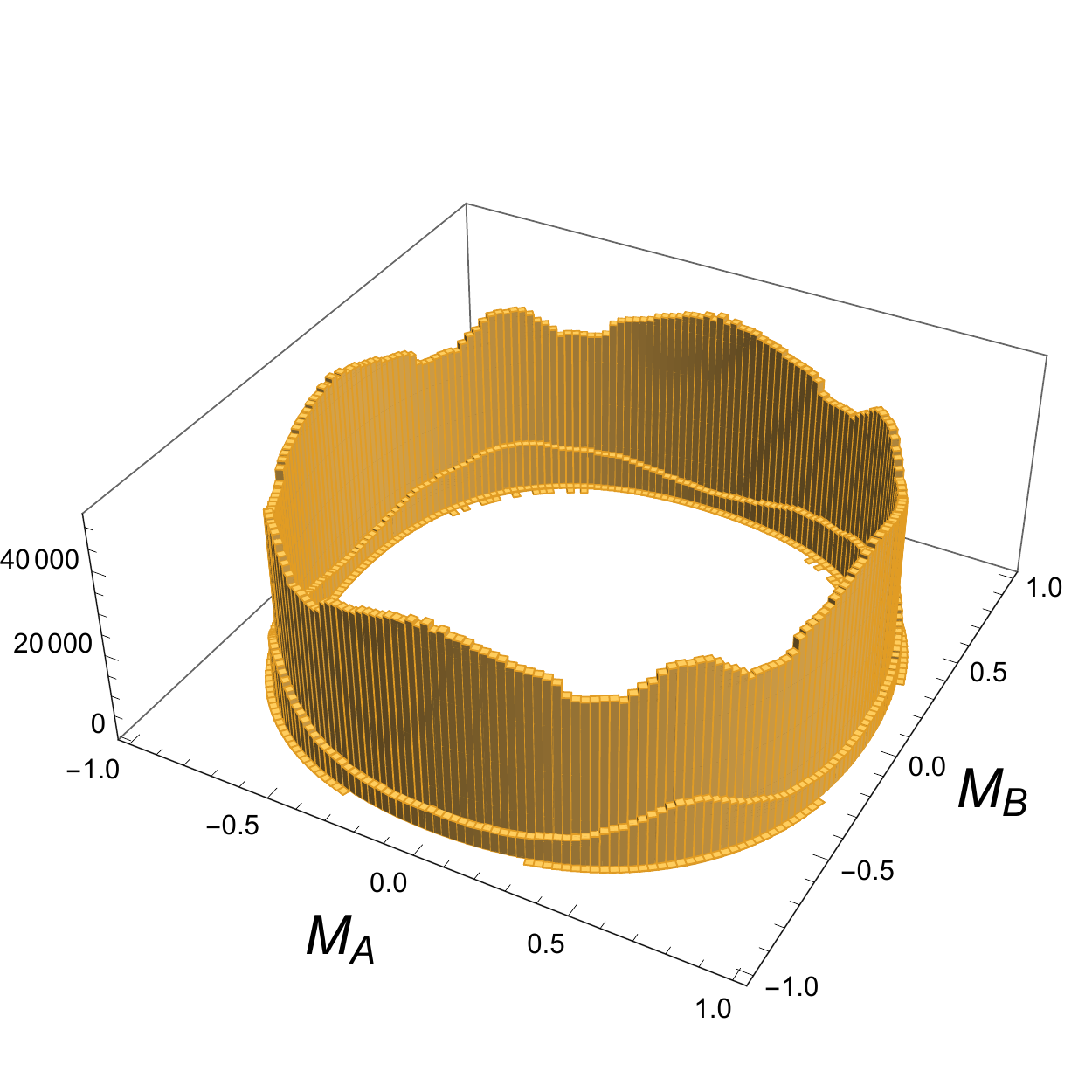}
\includegraphics[width=0.325\textwidth]{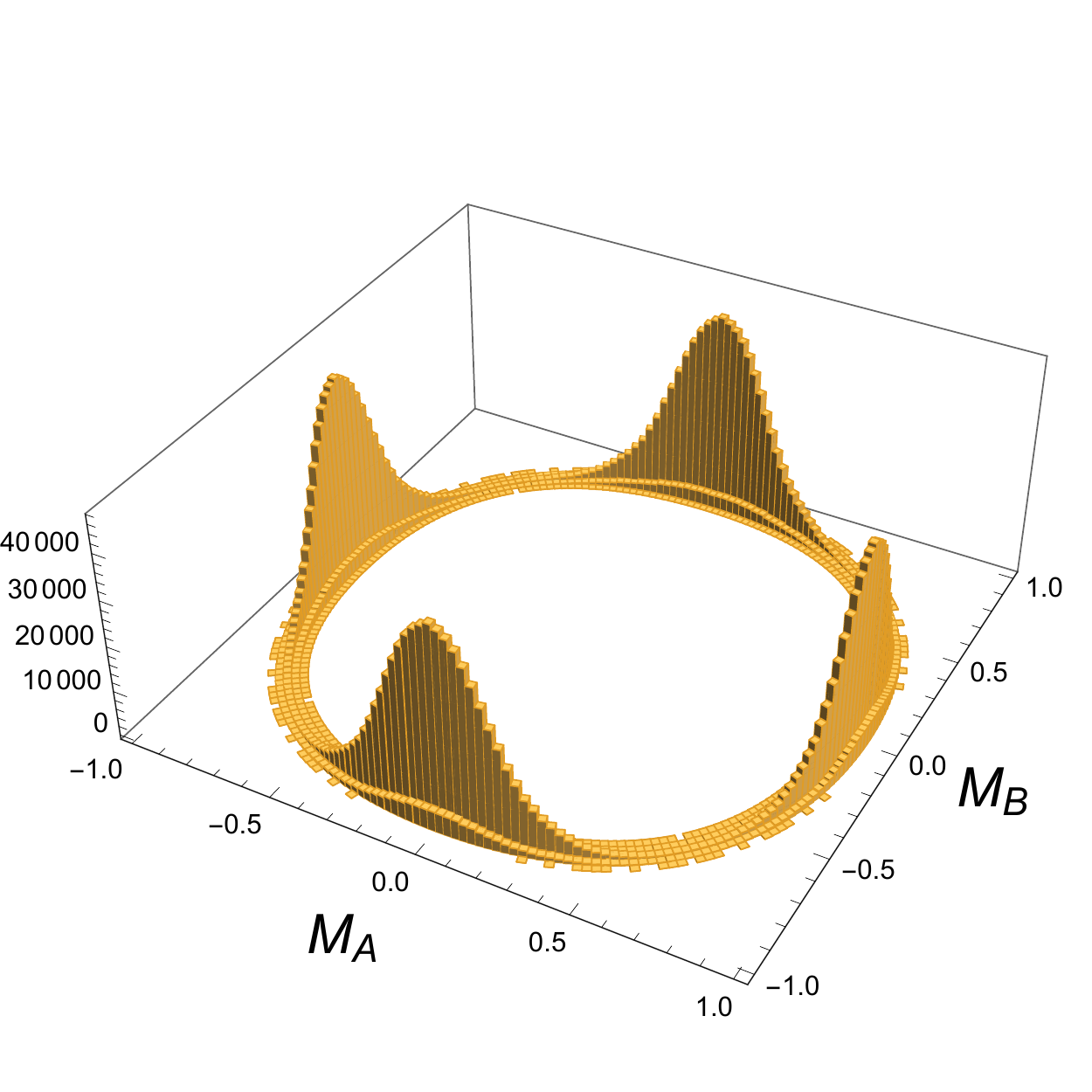}
\caption{[Color online] \textit{Order parameter distributions for the $U(1)$ 
quantum link model on the triangular lattice in the $(M_A,M_B)$ plane for 
$L = 64$ at $\lambda = -0.2156$ (left), $-0.2152 \approx \lambda_c$ (middle),
and $-0.2146$ (right).}}
\label{Fig2}
\end{figure}
The order parameter distributions over the $(M_A,M_B)$ plane are illustrated in 
Fig.\ref{Fig2}. There is a very weak first-order phase transition at 
$\lambda_c = - 0.215(1)$. In the phase at $\lambda < \lambda_c$ both order
parameters are non-zero $M_A, M_B \neq 0$, while for $\lambda > \lambda_c$ only
one sublattice orders. Both phases are characterized by the spontaneous
breakdown of lattice rotation invariance, and are qualitatively new
``nematic'' confined phases. Similarly, on the square lattice there are 
``crystalline'' confined phases in which lattice translation invariance is
spontaneously broken \cite{Ban13,Ban14}. Both on the triangular and on the 
square lattice, the phase transition that separates the two bulk confined phases
is characterized by a ring-shaped order parameter distribution indicating an 
emergent, approximate, global $SO(2)$ symmetry, which is spontaneously broken. 
The corresponding dual pseudo-Goldstone boson resembles an almost massless 
photon. However, since $(2+1)$-d $U(1)$ gauge theories are always confining, 
the pseudo-Goldstone boson is dual to a massive ``photon-ball''. Since the 
phase transitions are first order, one cannot take a continuum limit of these 
particular lattice models.

\begin{figure}[thb]
\centering \vskip-0.5cm
\includegraphics[width=0.24\textwidth]{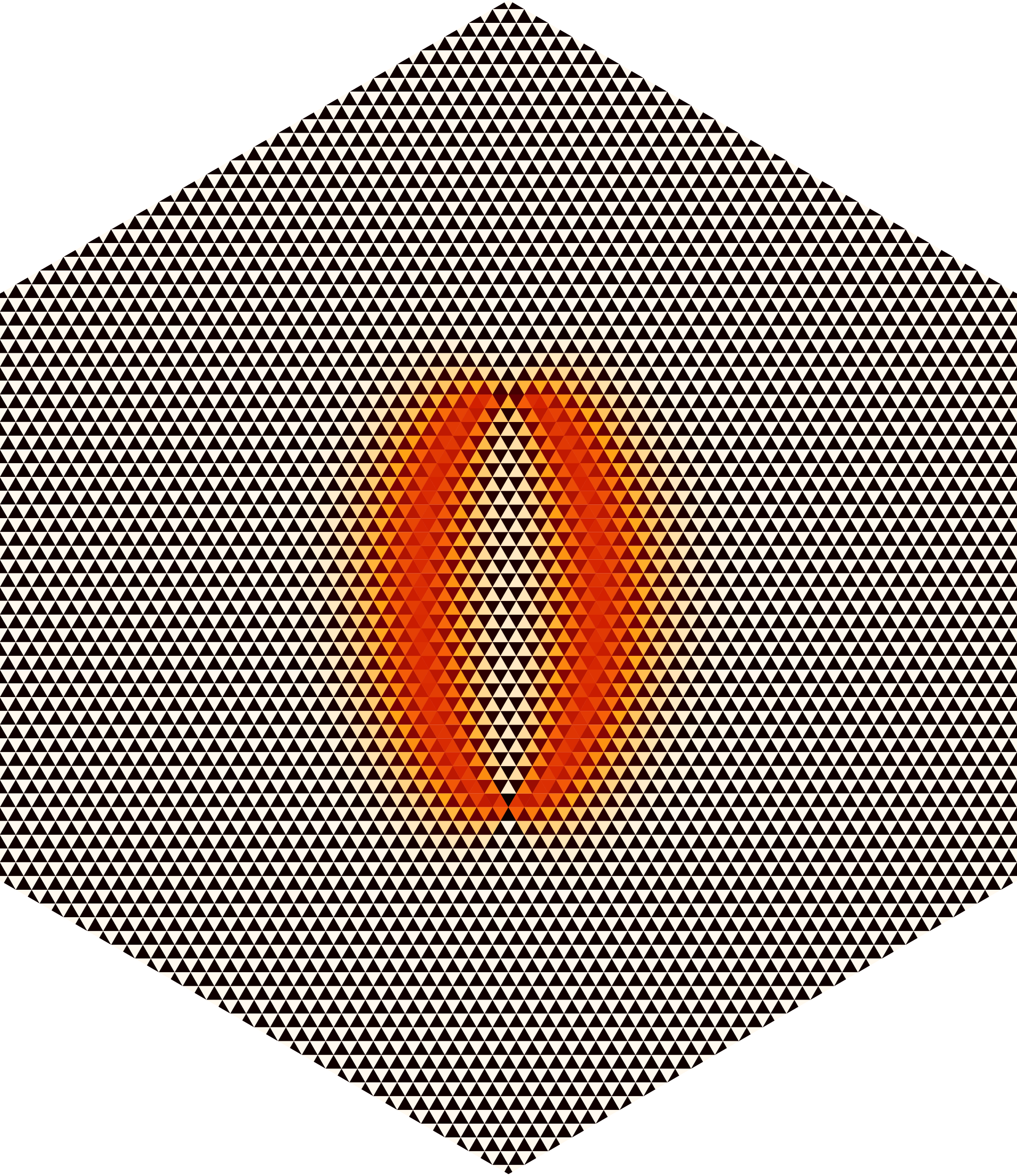}
\includegraphics[width=0.24\textwidth]{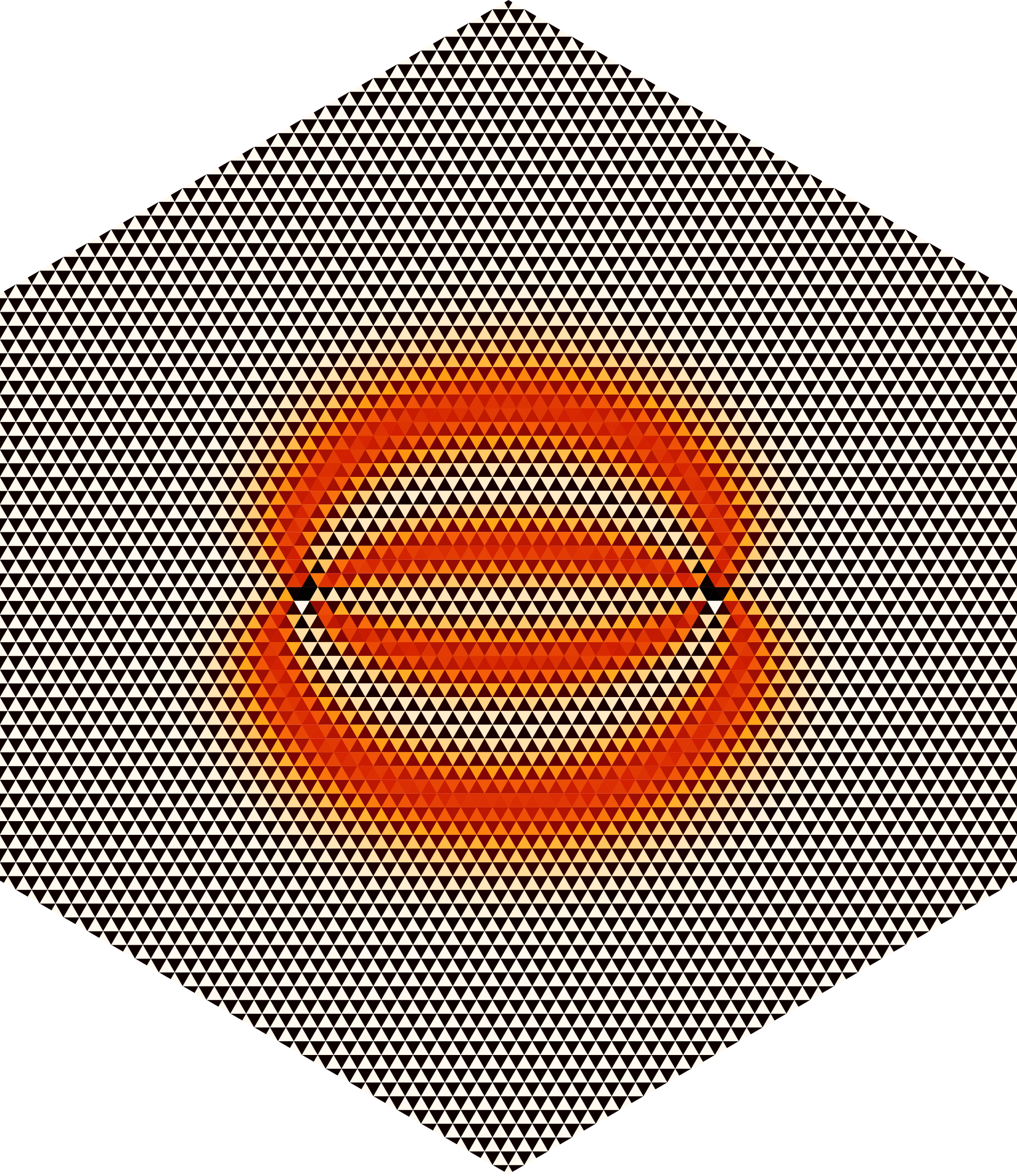}
\includegraphics[width=0.24\textwidth]{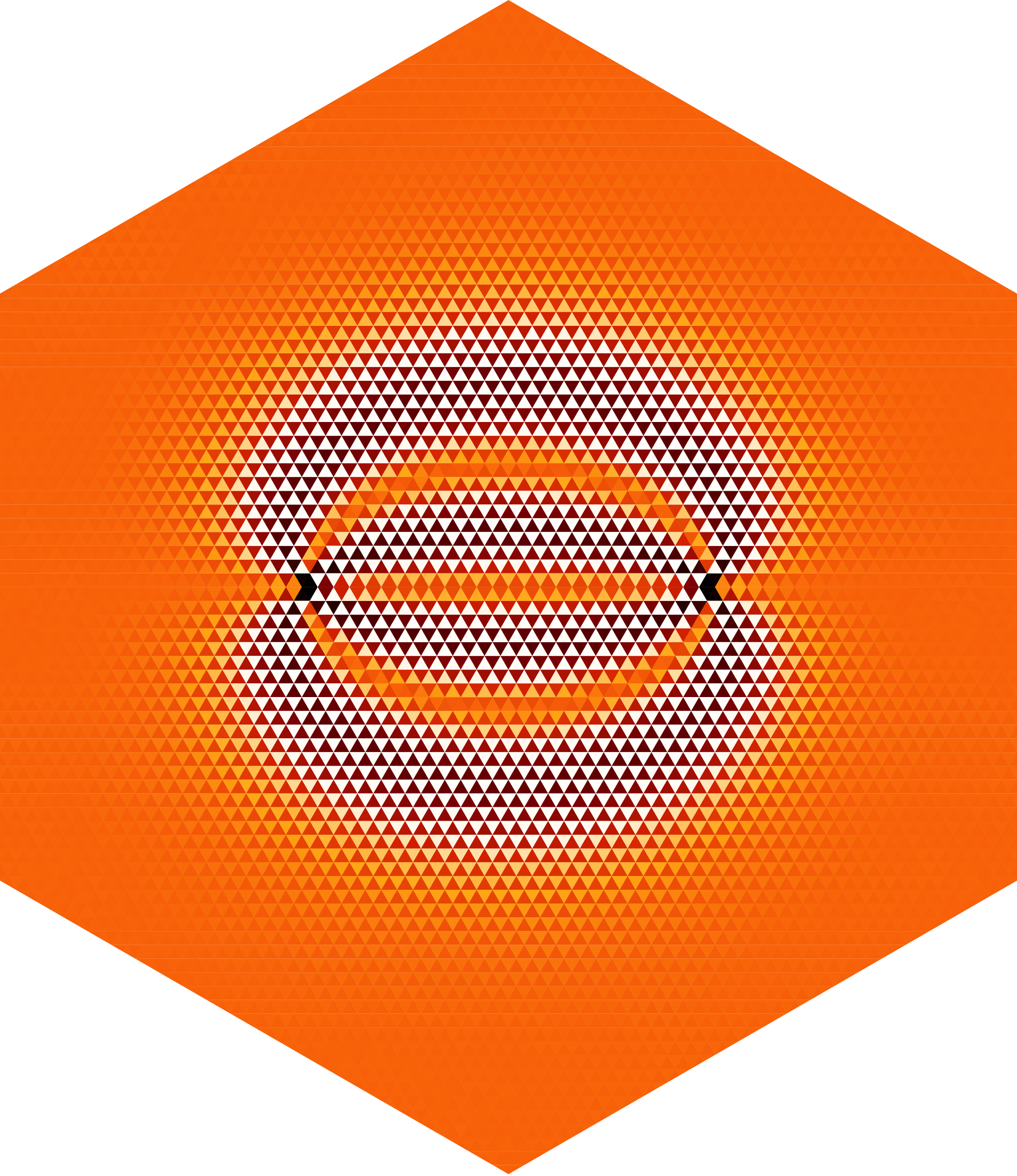}
\includegraphics[width=0.24\textwidth]{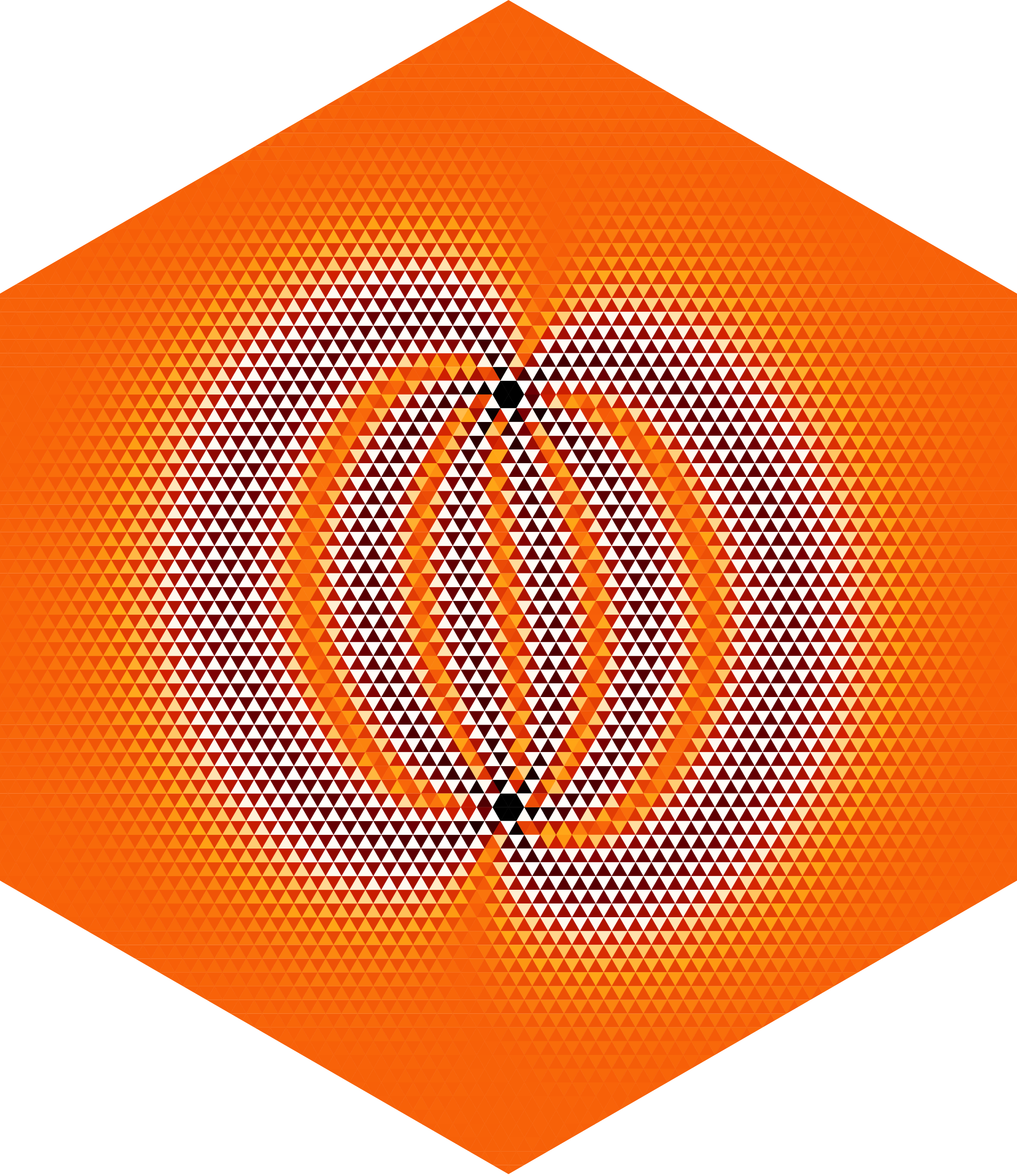}
\caption{[Color online] \textit{Energy distribution for the strings
connecting two charges $\pm 1$ at distance $r = 15 \sqrt{3}$ (a), and $\pm 2$ 
at $r = 26$ (b), with $\lambda = - 0.1 > \lambda_c$, as well as $\pm 3$ at 
distance $r = 15 \sqrt{3}$ (c), and $\pm 2$ at $r = 26$ (d), with 
$\lambda = -0.3 < \lambda_c$.}}
\label{Fig3}
\end{figure}
The energy density of the confining strings that connect external charges $Q_x$
and $Q_y = - Q_x$ located at distant lattice sites $x$ and $y$ are illustrated
in Fig.\ref{Fig3}. Remarkably, the string that connects the external charges 
fractionalizes into strands, each carrying fractional electric flux 
$\tfrac{1}{2}$. The strands are interfaces that separate the different bulk 
phases. Interestingly, the interior of the strands consists of the bulk phase 
that is realized on the other side of the phase transition.

\begin{figure}[tb]
\centering
\includegraphics[width=0.48\linewidth]{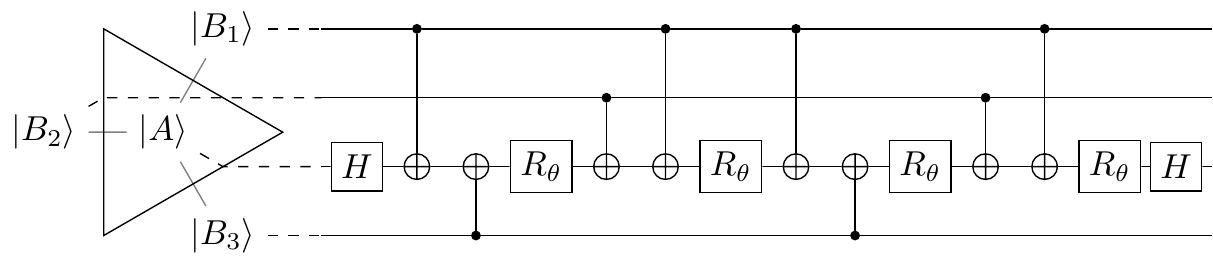}
\includegraphics[width=0.48\linewidth]{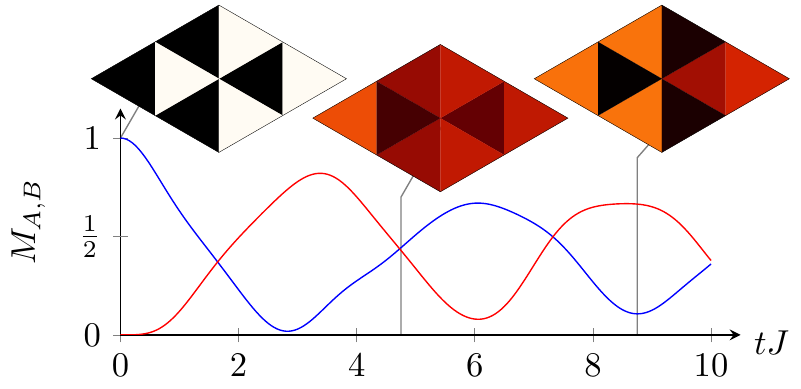}
\caption{\textit{Left: Circuit decomposition of the unitary time-evolution 
operator of a single triangular plaquette at $\lambda = 0$, over a discrete
time-step $\Delta t$, using two Hadamard gates $H$, four single qubit rotations 
$R_\theta$ with $\theta = - \Delta t J/2$ and eight CNOT gates. Right: 
Real-time evolution of the order parameters $M_A$ and $M_B$ on 8 triangular
plaquettes. The energy density is illustrated at three times, 
$t J = 0, 4.75, 8.75$ in the same way as in Fig.\ref{Fig3}.}}
\label{Fig4}
\end{figure}

The dual height representation has been used to implement the square lattice
$U(1)$ quantum link model on a configurable arrays of Rydberg atoms 
\cite{Cel20}. As illustrated in Fig.\ref{Fig4}, for the model on the triangular 
lattice, the use of the dual height variables gives rise to a particularly 
resource efficient encoding in a quantum circuit \cite{Ban21}. In this way the 
real-time dynamics of the confining strings is accessible to quantum 
simulations on near-term devices.

\subsection{D-Theory: Continuum Physics from Dimensional 
Reduction}

The $(2+1)$-d $U(1)$ quantum link models discussed before have first order
phase transitions and thus do not give rise to a continuum limit. In the Wilson
theory with its infinite-dimensional link Hilbert space, on the other hand,
a continuum limit is obtained at a second order phase transition that is reached
in the weak coupling limit $e \rightarrow 0$. Polyakov was first to argue that
$U(1)$ gauge theories in three space-time dimensions confine at all values of 
the gauge coupling \cite{Pol77}. This is due to the proliferation of magnetic 
monopoles, which are instantaneous events in 3-d space-time that play the role 
of instantons. G\"opfert and Mack \cite{Goe81} proved rigorously that the 
correlation length, which represents the inverse mass of a confined 
``photon-ball'', diverges as $\xi \sim \exp(c/e^2)$ in the weak coupling limit, 
thus showing that confinement persists at all couplings. Taking the continuum 
limit in the Wilson theory may not be the most practical approach when quantum 
simulations or computations shall be employed in order to investigate the 
real-time evolution.

Quantum link models approach the continuum limit in their own way, namely by 
the dimensional reduction of discrete variables, which is natural in the 
D-theory framework. In order to understand this way of taking the continuum 
limit, we start out with the theory in one more spatial dimension. Hence, we 
consider the $U(1)$ quantum link model on a 3-d spatial lattice. In a 4-d 
space-time, monopoles are no longer event-like but represent particles that 
travel along their worldlines. When monopoles condense, they lead to confinement
(just as in the lower-dimensional theory). However, in a 4-d space-time $U(1)$ 
gauge theories also possess Coulomb phases with a massless unconfined photon. 
In the Wilson formulation of $U(1)$ gauge theory on a 4-d space-time lattice, 
the Coulomb phase is separated from the confined phase by a weak first order 
quantum phase transition in the bare gauge coupling $e$.

For concreteness, let us consider the $U(1)$ quantum link model on a 3-d cubic 
spatial lattice. It is plausible that this model exists in a $(3+1)$-d Coulomb 
phase even when it is realized in the extreme quantum limit with quantum spins 
$\tfrac{1}{2}$ on each link \cite{Cha98}. A Coulomb phase is characterized by an
infinite correlation length $\xi = \infty$ associated with the massless photon. 
It is interesting to ask what happens when one compactifies one of the spatial 
dimensions to a finite extent $L'$. If $\xi$ would remain infinite, the Coulomb 
phase would persist even in $(2+1)$-d. However, $(2+1)$-d $U(1)$ gauge theories 
are known to be always confining. The effective gauge coupling $e$ of the
dimensionally reduced $(2+1)$-d theory is related to the gauge coupling $e'$ of
the $(3+1)$-d theory by $1/e^2 = L'/{e'}^2$, which implies
$\xi \sim \exp(c/e^2) = \exp(c L'/{e'}^2) \gg L'$. Interestingly, with 
increasing extent $L'$ of the finite spatial dimension, the correlation length 
$\xi$ increases exponentially, and becomes much larger than $L'$ itself. As a 
result, the theory undergoes dimensional reduction from $(3+1)$-d to $(2+1)$-d. 
The dimensional reduction of discrete variables is characteristic of D-theory, 
which provides a natural way to take the continuum limit in quantum link 
models. Unlike in the Wilson framework, where one tunes the value of the 
coupling $e$, in D-theory one just moderately increases the extent of an extra 
dimension. In practice, the extent of the extra dimension is just a few
lattice spacings, because the correlation length $\xi$ responds exponentially
to $L'$. In this way, one piles up discrete quantum link variables in an extra
dimension, in order to provide the minimal number of degrees of freedom that
are necessary to approach the continuum limit in a resource efficient manner.

\section{Non-Abelian Hamiltonian Lattice Gauge 
Theories}

Non-Abelian gauge theories play a central role in the standard model of 
particle physics. In particular, the strong interaction between quarks is 
mediated by the $SU(3)$ gluon gauge field of QCD. Non-Abelian gauge theories 
are also important in quantum information science, in particular, in the 
context of topological quantum computation, which is based on $(2+1)$-d 
Chern-Simons gauge theories \cite{Nay08}. In this section we discuss 
non-Abelian lattice gauge theories in the Hamiltonian formulation, first with 
Wilson's lattice gauge theory and then using quantum link models. Again, via 
the dimensional reduction of discrete variables, D-theory offers a natural way 
of approaching the continuum limit.

\subsection{Analog ``Particles'' Moving in the Group Manifold 
$SU(2) = S^3$ }

Let us consider the quantum mechanical analog ``particle'' for a Wilson-type 
parallel transporter in an $SU(2)$ lattice gauge theory. The corresponding
group manifold is the sphere $S^3$. Consequently, the position of the analog
``particle'' is described by an $SU(2)$ matrix
\begin{equation}
U = \cos\alpha + i \sin\alpha \vec e_\alpha \cdot \vec\sigma \ , \quad 
\vec e_\alpha = (\sin\theta \cos\varphi,\sin\theta \sin\varphi,\cos\theta) \ .
\end{equation}
Since $SU(2)$ is non-Abelian, we distinguish transformations that multiply
$U$ from the left and from the right. The corresponding 
$SU(2)_L \times SU(2)_R$ algebra is generated by
\begin{eqnarray}
&&\hskip-1.3cm \vec L = \frac{1}{2}(\vec J - \vec K), \
\vec R = \frac{1}{2}(\vec J + \vec K), \ J_\pm = \exp(\pm i \varphi)
\left(\pm \ \p_\theta + i \cot\theta \ \p_\varphi\right), \
J_3 =- i \p_\varphi, \nonumber \\ 
&&\hskip-1.3cm K_\pm = \exp(\pm i \varphi) \left(i \sin\theta \ \p_\alpha + 
i \cot\alpha \cos\theta \ \p_\theta \mp 
\frac{\cot\alpha}{\sin\theta} \p_\varphi\right), \
K_3 = i \left(\cos\theta \ \p_\alpha - \cot\alpha \sin\theta \ \p_\theta\right),
\end{eqnarray}
which obey the commutation relations $[\vec R,U] = U \vec \sigma$, 
$[\vec L,U] = - \vec \sigma U$. The kinetic energy of the analog ``particle'' 
corresponds to the Laplace-Beltrami operator (the Laplacian) of the group 
manifold, which together with its energy spectrum takes the form
\begin{equation}
T = \frac{1}{2 I} \left(\vec J^{\, 2} + \vec K^{\, 2}\right) =
\frac{1}{I} \left(\vec R^{\, 2} + \vec L^{\, 2}\right) \ , \
E_l = \frac{j_L(j_L + 1) + j_R(j_R + 1)}{I} = \frac{l(l + 2)}{2 I} \ .
\end{equation}
Here $j_L = j_R$ with $l = j_L + j_R \in \{0,1,2,...\}$ and each state
is $(2 j_L + 1)(2 j_R + 1) = (l+1)^2$-fold degenerate. Since the number of 
eigenstates is infinite, the corresponding Hilbert space is again
infinite-dimensional.

Just as in the Abelian case, we again consider three ``particles'' moving 
in the group manifold, associating the ``particles'' with the links 12, 23, 
and 31 of a triangular plaquette. The corresponding Hamiltonian now takes the
form
\begin{eqnarray}
H&=&T_{12} + T_{23} + T_{31} + V_{123} \nonumber \\
&=&e^2 \left(\vec R_{12}^{\, 2} + \vec L_{12}^{\, 2} +
\vec R_{23}^{\, 2} + \vec L_{23}^{\, 2} + 
\vec R_{31}^{\, 2} + \vec L_{31}^{\, 2} \right) -
\frac{1}{4 e^2}\mbox{Tr}(U_{12} U_{23} U_{31} + 
U_{31}^\dagger U_{23}^\dagger U_{12}^\dagger) \ .
\end{eqnarray}
The Hamiltonian commutes with the three infinitesimal gauge generators 
\begin{equation}
\vec G_1 = \vec L_{12} + \vec R_{31}, \
\vec G_2 = \vec L_{23} + \vec R_{12}, \
\vec G_3 = \vec L_{31} + \vec R_{23}, \
[H,\vec G_1] = [H,\vec G_2] = [H,\vec G_3] = 0.
\end{equation}

Again, by introducing an entire triangular lattice with an analog ``particle''
associated with each link, we now construct an $SU(2)$ lattice gauge theory
\begin{equation}
H = e^2 \sum_{\langle xy \rangle} \left(\vec R_{xy}^{\, 2} + \vec L_{xy}^{\, 2}\right)
- \frac{1}{4 e^2} \sum_{\langle xyz \rangle} \mbox{Tr}(U_{12} U_{23} U_{31} + 
U_{31}^\dagger U_{23}^\dagger U_{12}^\dagger) \ .
\label{SU2Hamiltonian} 
\end{equation}
In this case, the gauge generators associated with the lattice sites $x$ obey
\begin{equation}
\vec G_x = \sum_i (\vec L_{x,x+\hat i} + \vec R_{x-\hat i,x}) \ , \quad
[H,\vec G_x] = 0 \ , \quad [G_x^a,G_y^b] = i \delta_{xy} \epsilon_{abc} G_x^c \ .
\end{equation}
The non-Abelian Gauss law takes the form $\vec G_x|\Psi\rangle = 0$. Local
violations of Gauss' law manifest themselves as external static non-Abelian
gauge charges, which are characterized by an $SU(2)$ representation 
$Q \in \{0,\tfrac{1}{2},1,\tfrac{3}{2},\dots\}$ and one of the $2Q + 1$ 
corresponding values $Q^3 \in \{-Q,-Q+1,\dots,Q-1,Q\}$, such that
$\vec G_x^{\, 2}|\Psi,Q,Q^3\rangle = Q(Q+1) |\Psi,Q,Q^3\rangle$,
$G_x^3|\Psi,Q,Q^3\rangle = Q^3 |\Psi,Q,Q^3\rangle$.

\subsection{Quantum Links as Building Blocks of Non-Abelian 
Gauge Theories}

We will now replace Wilson-type parallel transporters by non-Abelian quantum
links, in order to be able to address the gauge dynamics in a finite-dimensional
Hilbert space per link. A Wilson-type parallel transporter $U$ is a matrix that
takes values in the gauge group. Obviously, its matrix elements $U^{ij} \in \C$
commute with each other, $[U^{ij},U^{kl}] = 0$. When one insists (unnecessarily)
on this property, the commutation relations can only be realized in an 
infinite-dimensional Hilbert space. Just like Wilson-type parallel 
transporters, non-Abelian quantum links are matrices. However, their matrix 
elements $U^{ij}$ are non-commuting operators, such that $[U^{ij},U^{kl}] \neq 0$.

In a non-Abelian quantum link model, there is a link-based embedding
algebra which contains the quantum link $U$ as well as the generators of gauge
transformations $L^a$ and $R^a$ associated with the left and right end of a 
link. In addition, there may be an Abelian generator $E$. These generators
obey the same commutation relations as in the Wilson theory  
\begin{eqnarray}
&&[L^a,L^b] = i f_{abc} L^c, \ [R^a,R^b] = i f_{abc} R^c \ , \ 
[L^a,R^b] = [L^a,E] = [R^a,E] = 0 \ ,
\nonumber \\
&&[L^a,U] = - \lambda^a U, \ [R^a,U] = U \lambda^a, \ [E,U] = U \ ,
\end{eqnarray}
except that the elements of a quantum link matrix do not commute. Here the
$\lambda^a$ are generators of the gauge Lie algebra which satisfy
$[\lambda^a,\lambda^b] = i f_{abc} \lambda^c$. Depending on the gauge group, the 
commutation relations $[U^{ij},U^{kl}] \neq 0$ are such that the corresponding 
embedding algebra closes. In a $U(N)$ or $SU(N)$ gauge theory, an $N \times N$ 
quantum link matrix is built from $2 N^2$ Hermitean operators, which replace 
the real and imaginary parts of the complex-valued matrix elements 
$U^{ij} \in \C$ of a parallel transporter in the Wilson framework \cite{Bro99}. 
Together with the $2 (N^2 - 1)$ generators $L^a$ and $R^a$, and an Abelian 
generator $E$, for a $U(N)$ gauge theory this yields
\begin{equation}
U(N): \ U^{ij}, \ L^a, \ R^a, E, \ 2 N^2 + 2 (N^2 - 1) + 1 = 4 N^2 - 1 \
SU(2N) \ \mbox{generators} \ .
\end{equation}
For a $U(1)$ gauge theory the embedding algebra $SU(2N)$ reduces to $SU(2)$.
For an $SO(N)$ (or more precisely $\mbox{Spin}(N)$) gauge group, the quantum
link has $N^2$ elements that replace the real-valued matrix elements 
$U^{ij} \in \R$ of the Wilson theory \cite{Bro04}. In addition, there are 
$2 \tfrac{N(N-1)}{2}$ generators $L^a$ and $R^a$, leading to the embedding
algebra $SO(2N)$
\begin{equation}
SO(N): \ U^{ij}, \ L^a, \ R^a, \ N^2 + 2 \frac{N(N-1)}{2} = N(2N - 1) \
SO(2N) \ \mbox{generators} \ .
\end{equation}
Finally for the gauge group $Sp(N)$ there are $4 N^2$ Hermitean operators
describing the $2N \times 2N$ quantum link matrix and $2N(2N+1)$ generators
$L^a$ and $R^a$, leading to the embedding algebra $Sp(2N)$ \cite{Bro04}
\begin{equation}
Sp(N): \ U^{ij}, \ L^a, \ R^a, \ 4N^2 + 2N(2N+1) = 2N(4N + 1) \
Sp(2N) \ \mbox{generators} \ . 
\end{equation}
Since $SU(2) = SO(3) = Sp(1)$, an $SU(2)$ gauge theory can be realized with the
embedding algebras $SU(4) = SO(6)$ or $Sp(2) = SO(5)$, leading to the simplest 
$SU(2)$ quantum link model.

\subsection{The $SU(2)$ Quantum Link Model on a 
Honeycomb Lattice}

\begin{figure}[tbp] 
\includegraphics[width=0.2\textwidth]{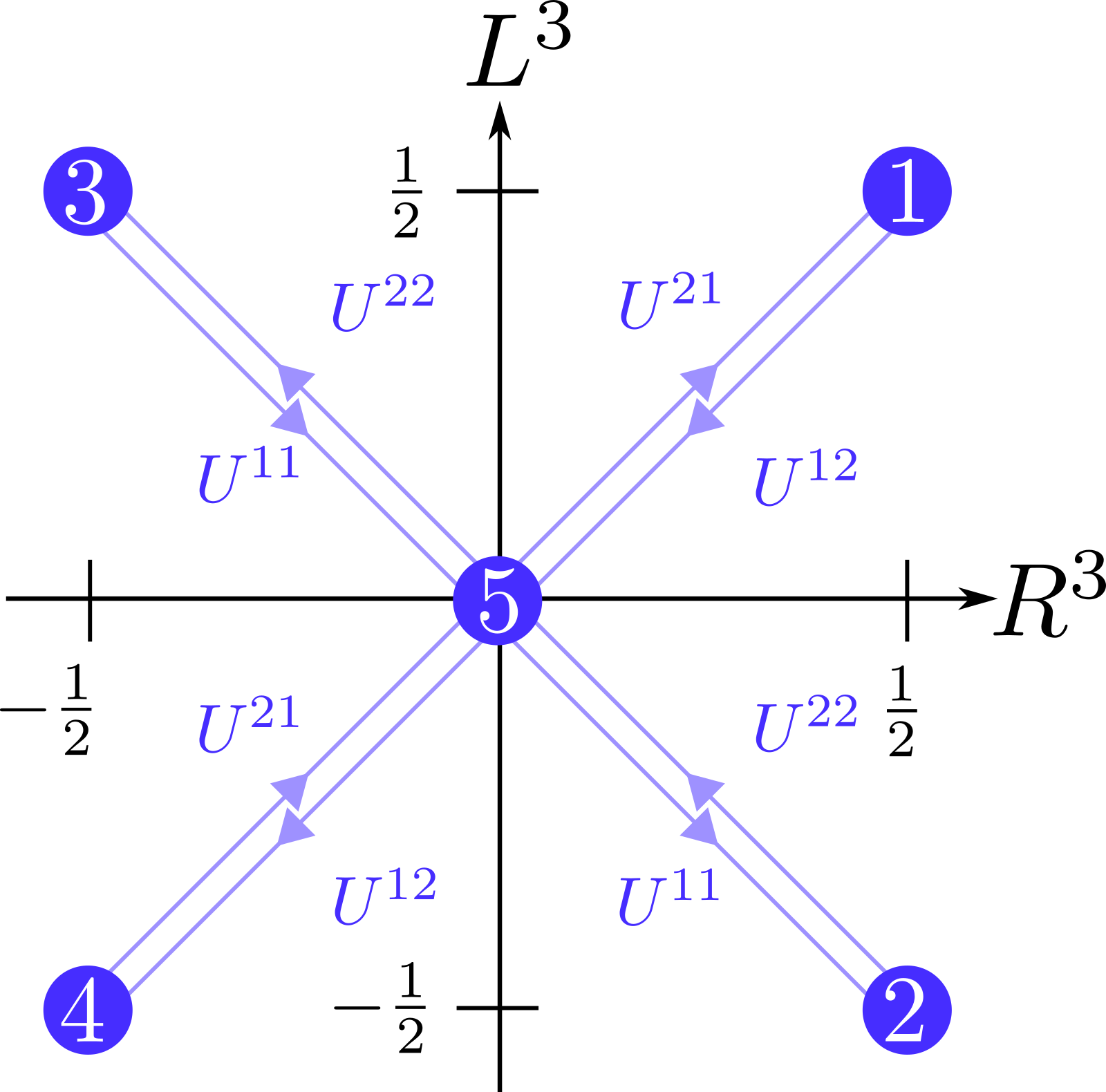} \hskip0.4cm 
\includegraphics[width=0.2\textwidth]{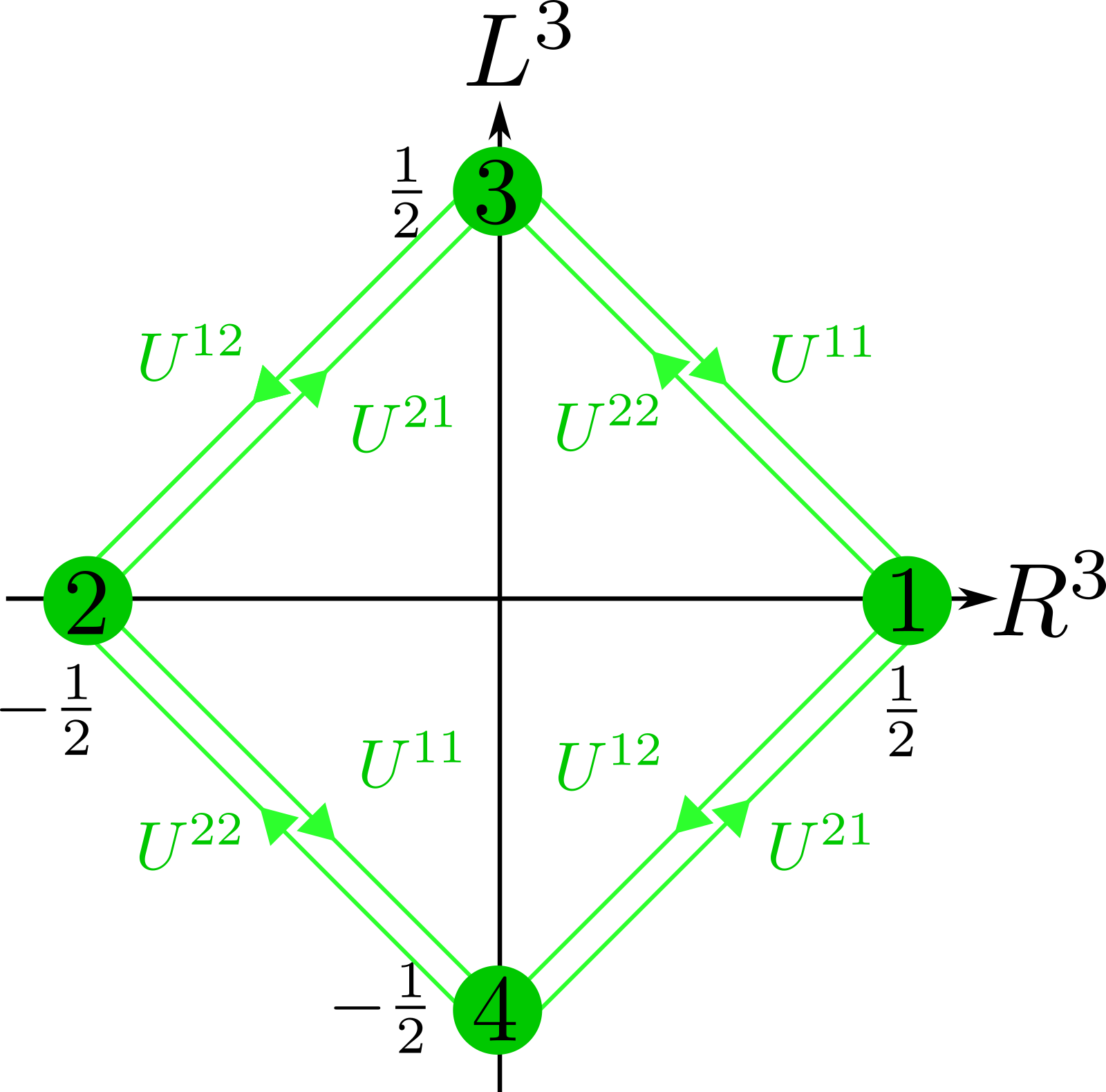} \hskip0.4cm
\includegraphics[width=0.45\textwidth]{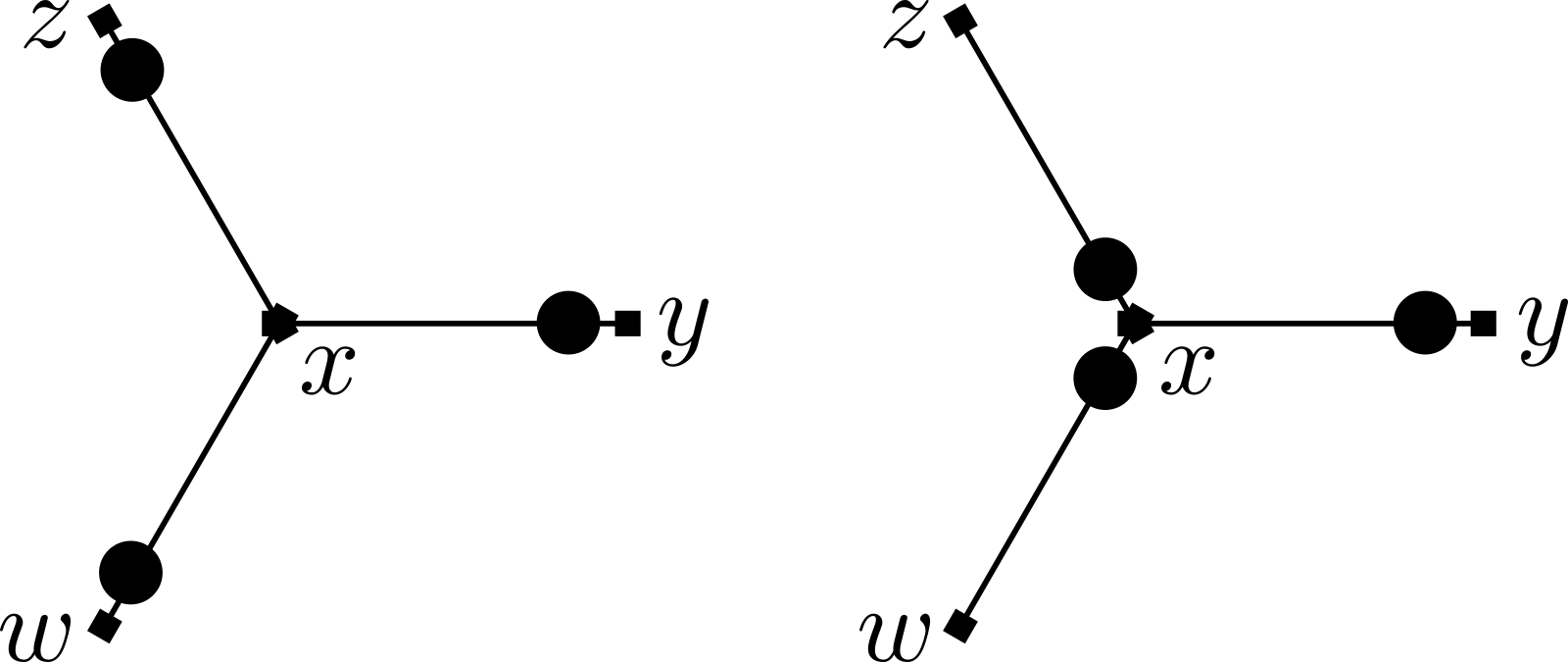} \\
\includegraphics[width=0.2\textwidth]{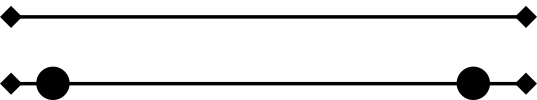} \hskip0.4cm 
\includegraphics[width=0.2\textwidth]{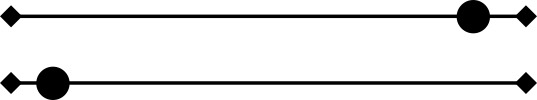}
\caption{[Color online] \textit{Left: Weight diagram of the 5-d vector
representation of $SO(5)$ above the corresponding link states. A black dot
symbolizes a spin $\tfrac{1}{2}$ attached to an end of a link. Middle: Weight 
diagram of the 4-d spinor representation of $SO(5)$ above the corresponding 
link states. Right: Two possible realizations of the Gauss law.}}
\label{Fig5}
\end{figure}

As a simple example of a non-Abelian quantum link model, let us consider the
$SU(2)$ quantum link model on a honeycomb lattice \cite{Ban18}. The simplest 
representations of the embedding algebra are the 5-d vector and the 4-d spinor 
representation of $SO(5)$, whose weight diagrams are illustrated in 
Fig.\ref{Fig5}. Under the $SU(2)_L \times SU(2)_R$ gauge transformations
at the left and right end of a link, they decompose as
\begin{equation}
\{5\} = \{1,1\} + \{2,2\} \ , \quad \{4\} = \{2,1\} + \{1,2\} \ . 
\end{equation}

\begin{figure}[tbp]
\includegraphics[width=0.48\textwidth]{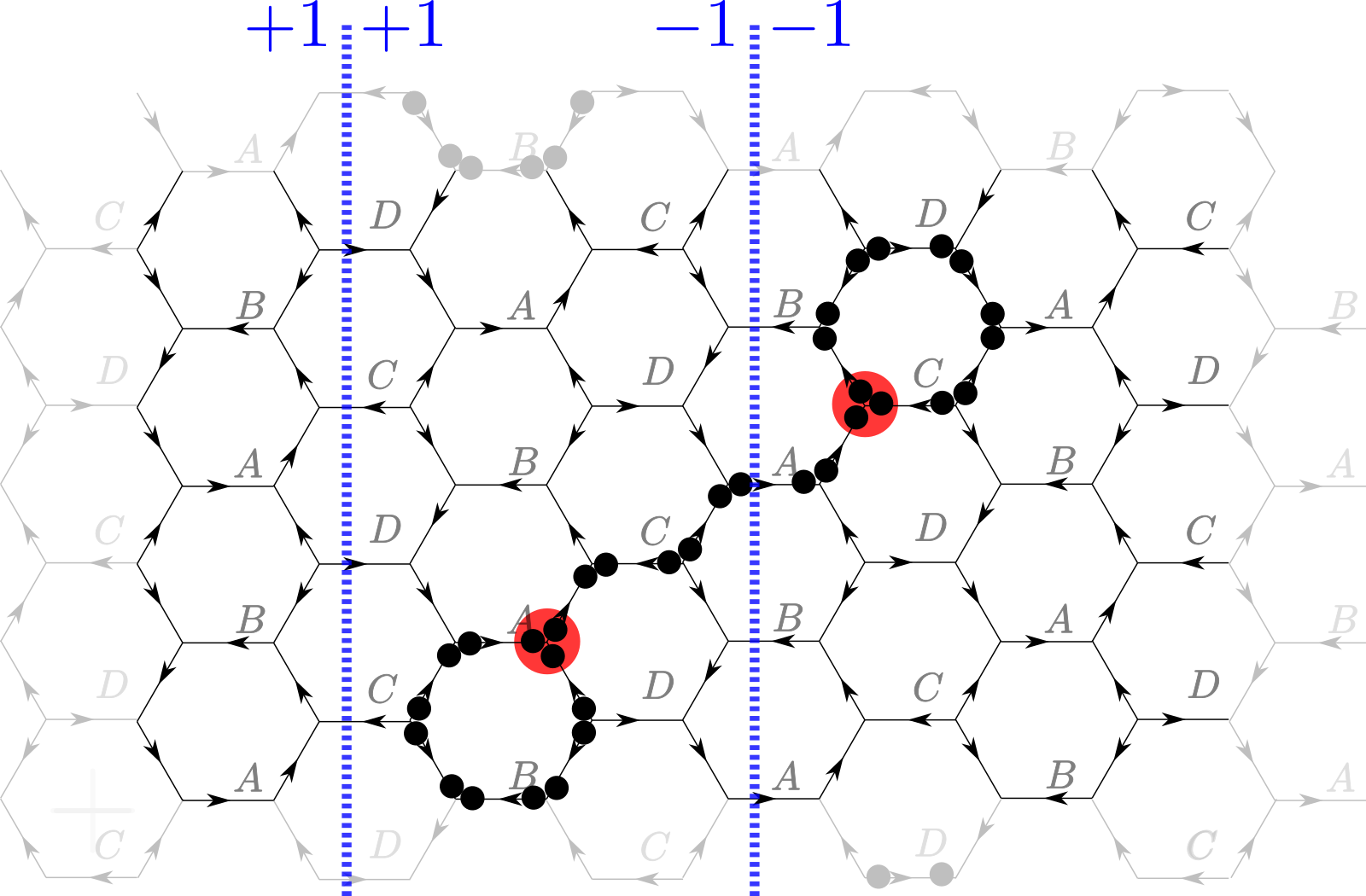} \hskip0.4cm
\includegraphics[width=0.48\textwidth]{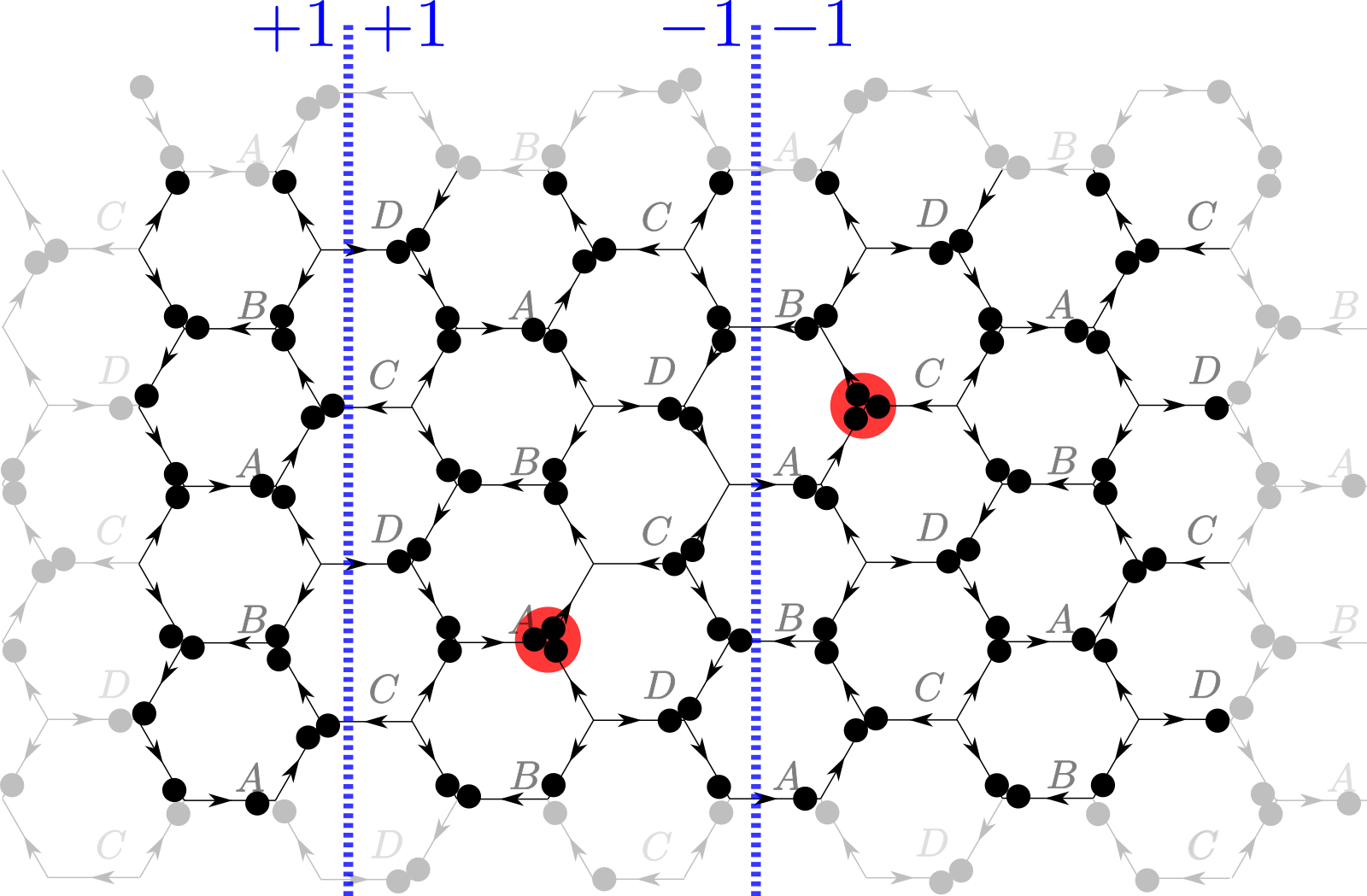}
\caption{[Color online] \textit{Flux configuration connecting two external
charges $Q = \tfrac{3}{2}$ in the $SU(2)$ quantum link model on the honeycomb
lattice for the 5-d vector (left) and 4-d spinor representation of $SO(5)$
(right).}}
\label{Fig6}
\end{figure}

The vector representation $\{5\}$ transforms trivially under the center $\Z(2)$ 
of the universal covering group $Spin(5)$ of the embedding algebra $SO(5)$. It
contains the state $\{1,1\}$ which corresponds to vanishing flux, i.e.\
$j_L = j_R = 0$, as well as four states $\{2,2\}$ with 
$j_L = j_R = \tfrac{1}{2}$. This resembles a truncation of the Wilson theory,
which is characterized by 
$j_L = j_R \in \{0,\tfrac{1}{2},1,\tfrac{3}{2},\dots\}$, to a 5-d Hilbert space
per link. A corresponding flux configuration connecting two external charges 
$Q = \tfrac{3}{2}$ is illustrated in Fig.\ref{Fig6}. The spinor 
representation $\{4\}$, on the other hand, transforms non-trivially under the 
center $\Z(2)$ and is characterized by $(j_L,j_R) = (\tfrac{1}{2},0)$ or
$(0,\tfrac{1}{2})$. Since now $j_L \neq j_R$, this model is qualitatively
different from the Wilson theory. There are two ways of satisfying Gauss' law
at a lattice site, which are again illustrated in Fig.\ref{Fig5}. A flux 
configuration connecting two external charges $Q = \tfrac{3}{2}$ that satisfies
the Gauss law is illustrated in Fig.\ref{Fig6}. The triangular lattice 
that is dual to the original hexagonal lattice can be divided into four 
sublattices with corresponding height variables that take values $\pm 1$. It is 
straightforward to extend the construction of a quantum circuit to the
$SU(2)$ quantum link model and to study the corresponding non-Abelian string
dynamics on a chip.

\subsection{$(1+1)$-d $\CP(N-1)$ Model from Dimensional
Reduction of a \\ $(2+1)$-d $SU(N)$ Quantum Spin Ladder}
 
Let us now consider the D-theory approach to the asymptotically free $(1+1)$-d
$\CP(N-1)$ models \cite{DAd78,Eic78}, which result from the dimensional 
reduction of discrete $SU(N)$ quantum spin variables \cite{Bea05}. Although 
these models have only a global $SU(N)$ symmetry, they share many features with 
non-Abelian gauge theories. In particular, they are asymptotically free, have a 
non-perturbatively generated mass gap, as well as a topological charge and 
$\theta$-vacuum states.

\begin{figure}[tbp]
\includegraphics[width=0.68\textwidth]{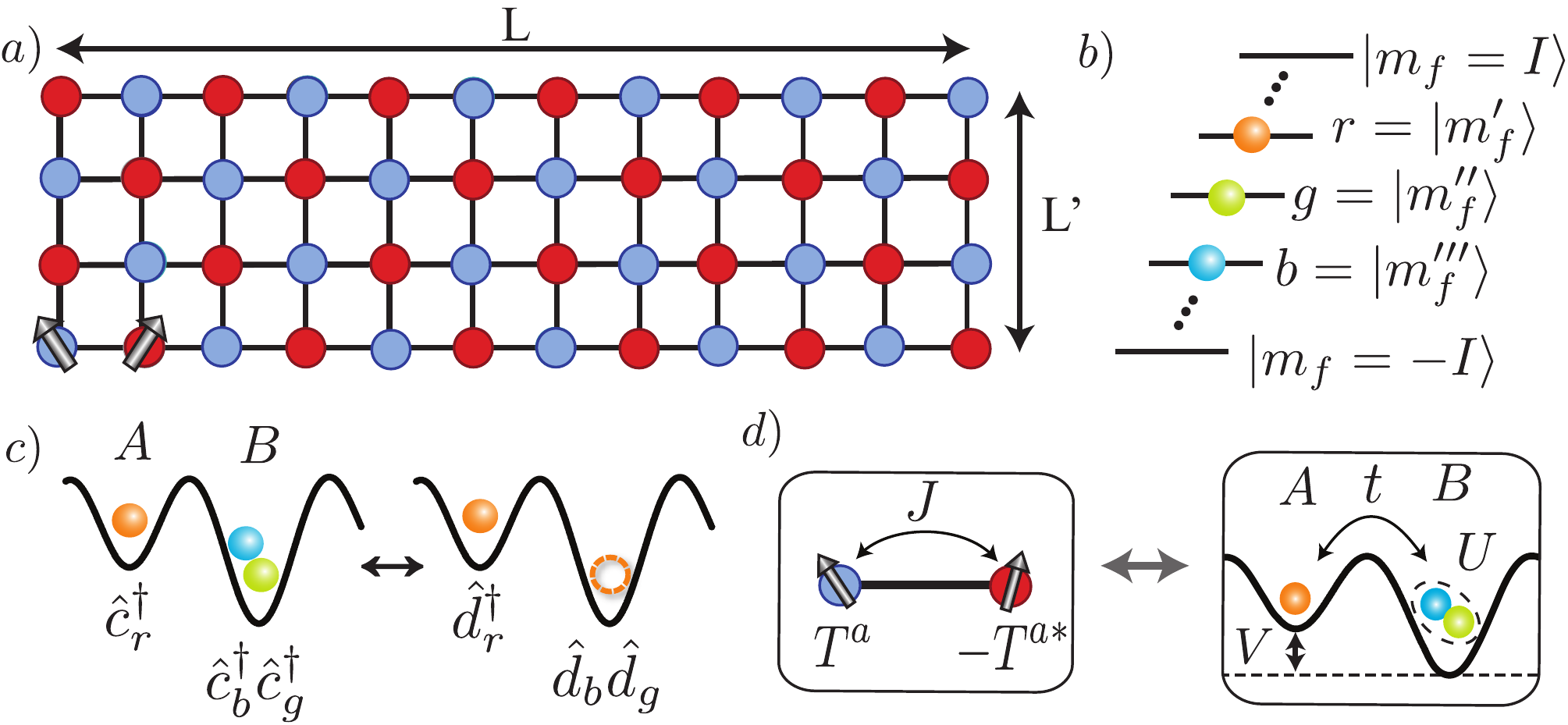} \hskip0.4cm
\includegraphics[width=0.3\textwidth]{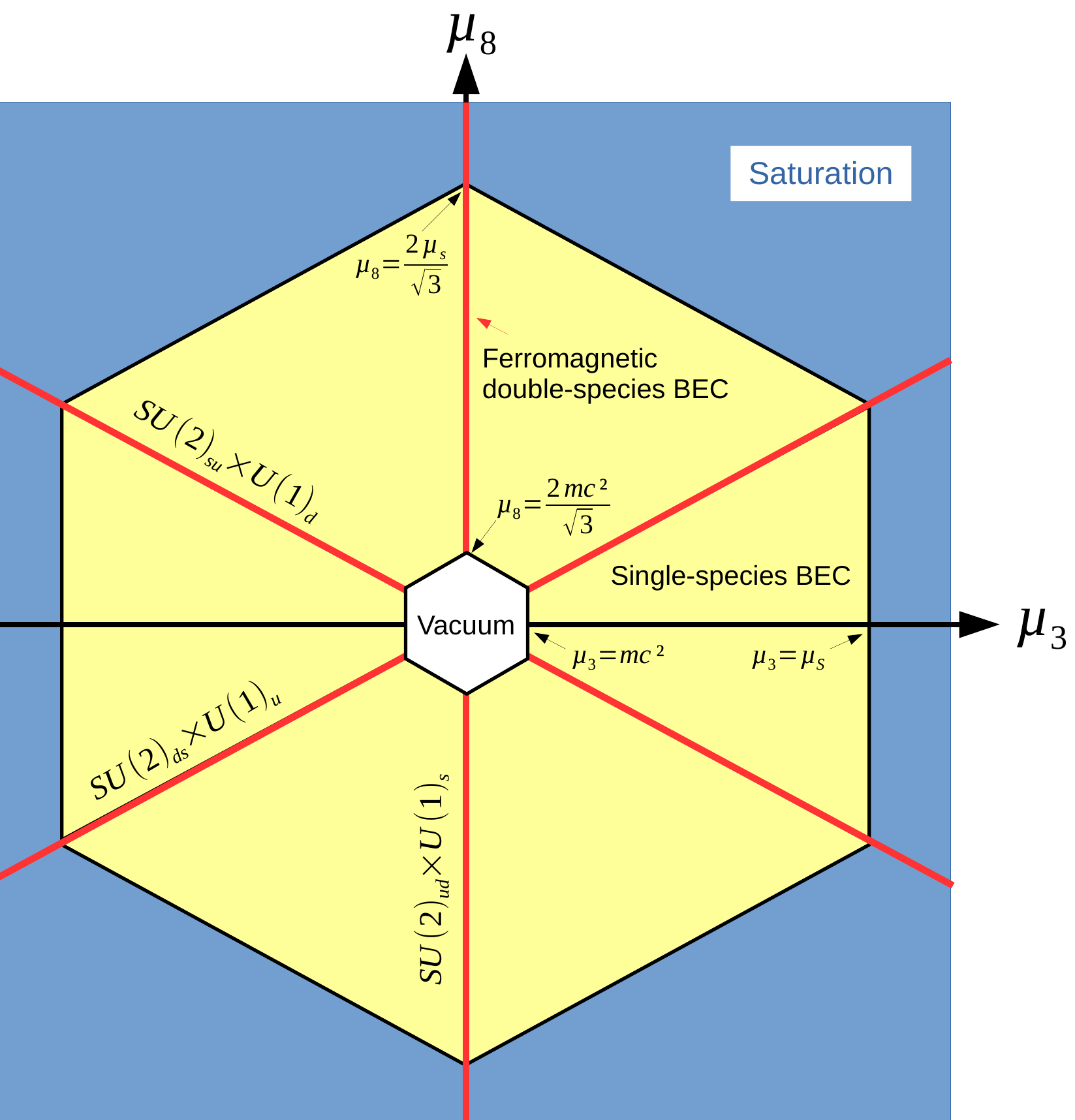}
\caption{[Color online] \textit{Left: $SU(3)$ quantum spin ladder embodied with
alkaline-earth atoms in an optical superlattice. a) Spin ladder with $SU(3)$ 
triplet spins $T^a$ on the even A and with anti-triplet spins $- T^{a*}$ on the 
odd B sublattice. b) Color degree of freedom encoded in the hyper-fine levels of
the nuclear spin. c) Hopping of atoms in an optical superlattice. Doubly
occupied sites encode anti-triplets. d) Triplet-anti-triplet spin interaction
realized by hopping and on-site repulsion of atoms. Right: Zero-temperature 
phase diagram of the $(1+1)$-d $\CP(2)$ model as a function of two chemical
potentials $\mu_3$ and $\mu_8$. The model possesses a rich ``condensed matter''
physics, with single- and double-species Bose-Einstein condensates, with and
without ferromagnetism.}}
\label{Fig7}
\end{figure}

Aiming at the $(1+1)$-d $\CP(2)$ model, let us now consider a 2-d bipartite 
square lattice of short even extent $L'$ in the 2-direction with open boundary 
conditions, as illustrated in Fig.\ref{Fig7}. In the continuum limit, the 
2-direction will disappear via dimensional reduction, while the 1-direction 
remains as the physical spatial dimension. We install $SU(3)$ triplet quantum 
spins $T^a_x$ on the even A sites and anti-triplet spins $-T^{a*}_y$ on the odd 
$B$ sites, in order to realize an anti-ferromagnetic spin ladder Hamiltonian 
that commutes with the total $SU(3)$ spin $T^a$
\begin{equation}
\label{Hamiltonian}
H = - J \sum_{\langle xy \rangle} T^a_x T^{a*}_y, \
[T_x^a,T_{x'}^b] = i \delta_{xx'} f_{abc} T_x^c, \
T^a = \sum_{x \in A} T^a_x - \sum_{y \in B} T^{a*}_y, \ [H,T^a] = 0.
\end{equation}
We couple a chemical potential to the conserved non-Abelian $SU(3)$ charge 
$T^a$ and obtain the grand canonical partition function
$Z = \mbox{Tr} \exp(- \beta (H - \mu_3 T^3 - \mu_8 T^8))$.

Let us first consider the system at zero temperature,
$\beta \rightarrow \infty$, in the infinite-volume limit, 
$L, L' \rightarrow \infty$. It turns out that the $SU(3)$ symmetry then breaks 
spontaneously to $U(2)$ \cite{Har03}, thus leading to $8 - 4 = 4$ massless 
Goldstone bosons, whose low-energy dynamics are described by an effective field 
theory in terms of $3 \times 3$ matrix fields $P(x)$ which take values in the 
coset space $SU(3)/U(2) = \CP(2)$, i.e.\ $P(x)^\dagger = P(x)$, $P(x)^2 = P(x)$, 
$\mbox{Tr} P(x) = 1$. When we make the extent $L'$ of the 2-direction 
finite, the Mermin-Wagner theorem implies that the continuous global $SU(3)$ 
symmetry can no longer break spontaneously. Hence, the Goldstone bosons 
pick up an exponentially small mass $m = 1/(\xi c)$. Their low-energy effective 
action takes the form
\begin{equation}
S[P] = \int_0^\beta dt \int_0^L dx_1 \int_0^{L'} dx_2 \ \rho_s \mbox{Tr}\left(
\partial_i P \partial_i P + \frac{1}{c^2} D_t P D_t P\right) \ , \ 
D_t P = \partial_t P - \mu_a [T^a,P] \ .
\end{equation}
Here $\rho_s$ is the spin stiffness and $c$ is the spinwave velocity. When 
$\xi \gg L'$ the field becomes $x_2$-independent and the system undergoes 
dimensional reduction from $(2+1)$-d to $(1+1)$-d with the dimensionless 
coupling constant $1/g^2 = L' \rho_s/c$, thus leading to the $\CP(2)$ model
action
\begin{eqnarray}
S[P] = \int_0^\beta dx_3 \int_0^L dx_1 \frac{1}{g^2} \mbox{Tr}\left[
\partial_1 P \partial_1 P + \frac{1}{c^2} D_t P D_t P\right] \ .
\end{eqnarray}
Due to asymptotic freedom of the $(1+1)$-d $\CP(2)$ model, the correlation 
length is exponentially large in $1/g^2$, i.e.\ $\xi \sim \exp(4 \pi/3 g^2) =
\xi \sim \exp(4 \pi L' \rho_s/3 c)$ (here $4 \pi/3$ is the 1-loop coefficient
of the $\beta$-function). This justifies the assumption that $\xi \gg L'$ 
already for moderately large values of $L'$. Dimensional reduction hence 
results as a consequence of asymptotic freedom.

The unconventional $(2+1)$-d $SU(N)$ quantum spin ladder regularization of the 
$(1+1)$-d $\CP(N-1)$ model makes their real-time dynamics accessible to quantum 
simulation experiments using ultracold alkaline-earth atoms 
($^{87}$Sr or $^{173}$Yb) in optical lattices \cite{Laf16}. Using a worm 
algorithm in Monte Carlo simulations on a classical computer the phase diagram 
of the model has been computed as a function of the chemical potentials 
$\mu_3$ and $\mu_8$ (cf.\ Fig.\ref{Fig7}) \cite{Eva18}. There are phases in 
which the massive bosons undergo single- or double-species Bose-Einstein 
condensation, the latter with ferromagnetism. It would be most interesting to 
perform quantum simulation experiments of the corresponding ``condensed matter 
physics'' of the $\CP(2)$ model.

\subsection{D-Theory: Continuum QCD from Dimensional 
Reduction}

Just like Abelian gauge fields in $(3+1)$-d, in $(4+1)$-d non-Abelian $SU(N)$
gauge fields can exist in a Coulomb phase with massless gauge bosons and hence
with an infinite correlation length $\xi$. The corresponding low-energy 
effective theory is a $(4+1)$-d Yang-Mills theory with the action
\begin{equation}
S[G_\mu] = \int dt \, d^3x \int_0^{L'} dx_4 \ 
\frac{1}{2 e^2} \mbox{Tr} \left( G_{\mu\nu} G_{\mu\nu} + 
\frac{1}{c^2} G_{\mu t} G_{\mu t}\right) \ , \ \mu, \nu \in \{1,2,3,4\} \ .
\end{equation}
When the extent $L'$ of the extra dimension becomes finite, due to confinement 
in $(3+1)$-d, $\xi$ cannot remain infinite. Assuming that $\xi \gg L'$, the
theory undergoes dimensional reduction from $4+1$ to 4 dimensions such that
\begin{equation}
S[G_\mu] \rightarrow \int dt \, d^3x \ 
\frac{1}{2 g^2} \mbox{Tr} \left(G_{ij} G_{ij} 
+ \frac{1}{c^2} G_{i t} G_{i t}\right), \ i, j \in \{1,2,3\}, \ 
\frac{1}{g^2} = \frac{L'}{e^2 c}, \ 
\frac{1}{m} \sim \exp\left(\frac{24 \pi^2 L'}{11 N e^2 c}\right).
\end{equation}
Again, the extent $L'$ of the extra dimension determines the asymptotically
free dimensionless gauge coupling $g$ (here the 1-loop coefficient of the 
$\beta$-function is $24 \pi^2/11 N$). Indeed $\xi \gg L'$ because, due to
asymptotic freedom, $\xi$ increases exponentially with $L'$ \cite{Cha97}.

It is very natural to incorporate Shamir's variant \cite{Sha93} of Kaplan's
domain wall fermions \cite{Kap92} in this $(4+1)$-d setup, which can be
regularized in the D-theory framework with $SU(N)$ quantum links \cite{Bro99}. 
The corresponding Hamiltonian is given by
\begin{eqnarray}
H&=&e^2 \sum_{x,\mu} \left[(R^a_{x,\mu})^2 + (L^a_{x,\mu})^2\right]
- \frac{1}{2 N e^2} \sum_{x,\mu \neq \nu} \mbox{Tr} [U_{x,\mu} U_{x+\hat\mu,\nu} 
U^\dagger_{x+\hat\nu,\mu} U^\dagger_{x,\nu}] \nonumber \\
&-&J' \sum_{x,\mu} \ [\mbox{det} U_{x,\mu} + \mbox{det} U^\dagger_{x,\mu}] +
\frac{1}{2} \sum_{x,\mu} \ [\Psi^\dagger_x \gamma_0 \gamma_\mu U_{x,\mu} 
\Psi_{x+\hat\mu}
- \Psi^\dagger_{x+\hat\mu} \gamma_0 \gamma_\mu U^\dagger_{x,\mu} \Psi_x] \nonumber \\
&+&M \sum_x \Psi^\dagger_x \gamma_0 \Psi_x +
\frac{r}{2} \sum_{x,\mu} \ [2 \Psi^\dagger_x \gamma_0 \Psi_x 
- \Psi^\dagger_x \gamma_0 U_{x,\mu} \Psi_{x+\hat\mu}
- \Psi^\dagger_{x+\hat\mu} \gamma_0 U^\dagger_{x,\mu} \Psi_x] \ . \nonumber
\end{eqnarray}
At finite extent $L'$, the domain wall fermions have a residual mass
$\mu = 2 M \exp(- M L')$. For a sufficiently large domain wall mass
$M > \frac{24 \pi^2}{(11 N - 2 N_f) e^2 c}$, the theory reaches the chiral
limit together with the continuum limit. The continuum limit is controlled by 
$\frac{1}{m} 
\propto \exp\left(\frac{24 \pi^2 \beta}{(11 N - 2 N_f) e^2 c}\right)$ in the
presence of $N_f$ flavors of quarks, which are described by the anti-commuting
fermion creation and annihilation operators $\Psi_x^\dagger$ and $\Psi_x$.
Ultracold alkaline-earth atoms in an optical superlattice can again be used
to embody $SU(N)$ quantum links in ultracold matter \cite{Ban13a}. This 
provides a concrete vision for how to ultimately quantum simulate QCD
\cite{Wie14}.

\section{Conclusion}

D-theory applied to quantum link models provides a formulation of gauge 
theories that allows their resource efficient implementation in quantum 
simulators or quantum computers. The continuum limit is reached naturally
(i.e.\ without fine-tuning) by a moderate increase of the size of an extra
spatial dimension. In the near future, close collaborations between theorists
and experimentalists hold the promise to realize many different aspects of 
strongly coupled gauge theories. Even if one works in a lower-dimensional 
space-time, with a smaller gauge group, a reduced matter content, or away from 
the continuum limit, once quantum simulations of gauge theories are realized 
experimentally, they become a very exciting subject in their own right. 
Exploring their real-time or finite-density dynamics, even at the qualitative 
level to which one will be limited without systematic error correction, is 
most interesting along the way towards ultimately quantum simulating QCD
\cite{Wie14}.

\enlargethispage{20pt}


\funding{The research of UJW is supported by the Schweizerischer Nationalfonds.}

\ack{I like to thank Steven Bass for inviting me to write this contribution.
I'm indebted to D.\ Banerjee, B.\ B.\ Beard, W.\ Bietenholz, M.\ B\"ogli, 
R.\ Brower, W.\ Evans, S.\ Caspar, S.\ Chandrasekharan, M.\ Dalmonte, 
F.-J.\ Jiang, U.\ Gerber, M.\ Hafezi, M.\ Hornung, C.\ Laflamme, D.\ Marcos, 
H.\ Mejia-Diaz, M.\ M\"uller, T.\ Z.\ Olesen, P.\ Orland, J.-H.\ Peng, 
M.\ Pepe, P.\ Rabl, E.\ Rico, S.\ Riederer, P.\ Stebler, M.\ Troyer, 
P.\ Widmer, and P.\ Zoller for very fruitful collaborations on the subjects 
discussed here.}

\end{document}